\documentstyle[11pt]{article}
\newcommand{\bra}{\langle}
\newcommand{\ket}{\rangle}
\newcommand{\R}{\mbox{\boldmath $ R $}}
\newcommand{\C}{\mbox{\boldmath $ C $}}
\newcommand{\Z}{\mbox{\boldmath $ Z $}}
\newcommand{\Rplus}{\mbox{\boldmath $ R $}_{\ge 0}}
\newcommand{\Ad}{\mbox{Ad}}
\newcommand{\g}{\mbox{\bf g}}
\newcommand{\Kernel}{\mbox{Ker} \,}
\newcommand{\Image}{\mbox{Im}}
\newcommand{\so}{\mbox{\bf so}(3)}

\newcommand{\vect}[1]{\mbox{\boldmath $ #1 $}}

\def\dfrac#1#2{\displaystyle\frac{#1}{#2}}

\newcommand{\mapright}[2]
{\mathop{\hbox to 1cm{\rightarrowfill}}
\limits^{\scriptstyle #1}_{\scriptstyle #2}}

\newcommand{\mapleft}[2]
{\mathop{\hbox to 1cm{\leftarrowfill}}
\limits^{\scriptstyle #1}_{\scriptstyle #2}}

\newcommand{\mapdown}[2]
{\Big \downarrow 
\llap {$\vcenter {\hbox{$\scriptstyle #1 \,$}}$ }
\rlap {$\vcenter {\hbox{$\scriptstyle #2   $}}$ }
}

\newcommand{\mapne}[1]
{\nearrow \llap{$\vcenter {\hbox{$\scriptstyle #1 \quad $}}$ }}

\newcommand{\isoarrow}
{\mathop{\hbox to 5mm{\rightarrowfill}}
\limits^{\scriptstyle \sim}}

\makeatletter
\@addtoreset{equation}{section}
\makeatother

\begin{document} 
\baselineskip 7mm
\vspace*{10mm}
\begin{center}
{\bfseries \Large
Reduction of Quantum Systems on Riemannian Manifolds
with Symmetry and Application to Molecular Mechanics}
\footnote{
Archive number math-ph/9907005.
Published in Journal of Mathematical Physics 41, 1814-1842 (2000).
}
\vspace{5mm} \\
Shogo Tanimura

{\it Department of Engineering Physics and Mechanics \\
Kyoto University, Kyoto 606-8501, Japan}

and 

Toshihiro Iwai

{\it Department of Applied Mathematics and Physics \\
Kyoto University, Kyoto 606-8501, Japan}
\vspace{4mm} \\
e-mail:
stani@yukawa.kyoto-u.ac.jp; iwai@amp.i.kyoto-u.ac.jp
\vspace{4mm} \\
Abstract

\begin{minipage}[t]{130mm}
This paper deals with a general method for the reduction of
quantum systems with symmetry.
For a Riemannian manifold $ M $
admitting a compact Lie group $ G $ as an isometry group,
the quotient space $ Q = M/G $ is not a smooth manifold in general
but stratified into a collection of smooth manifolds of
various dimensions.
If the action of the compact group $ G $ is free, $ M $ is made
into a principal fiber bundle with structure group $ G $.
In this case, reduced quantum systems are set up as
quantum systems on the associated vector bundles over $ Q = M/G $.
This idea of reduction fails, 
if the action of $ G $ on $ M $ is not free.
However, the Peter-Weyl theorem works well for reducing
quantum systems on $ M $.
When applied to the space of wave functions on $ M $,
the Peter-Weyl theorem provides the decomposition of
the space of wave functions into spaces of equivariant functions
on $ M $, which are interpreted as Hilbert spaces for
reduced quantum systems on $ Q $.
The concept of connection on a principal fiber bundle
is generalized to be defined well on the stratified manifold $ M $.
Then the reduced Laplacian is well defined as a self-adjoint
operator with the boundary conditions on singular sets of lower
dimensions.
Application to quantum molecular mechanics is also discussed in detail.
In fact, the reduction of quantum systems studied in this paper
stems from molecular mechanics.
If one wishes to consider the molecule which is allowed to lie
in a line when it is in motion, the reduction method presented
in this paper works well.
\end{minipage}
\end{center}
\vspace{4mm}
\\
PACS indices: 
02.20.-a, 02.40.-k, 31.15.-p, 33.20.Vq
%
\newpage
\baselineskip 5.8mm 
\section{Introduction}
Symmetry has always played an important role in mechanics.
When a Hamiltonian system admits a symmetry group,
the system reduces to a Hamiltonian system 
of less degrees of freedom.
Reduction of Hamiltonian systems with symmetry is established
by Marsden and Weinstein\cite{Marsden-Weinstein} 
and is explained in a textbook\cite{Abraham-Marsden}.
The reduction method has found a wide variety of applications.
In particular, the notion of moment map, which is a key word 
in the reduction method, has been introduced in differential geometry.

The reduction method is applicable to molecular mechanics, 
both classical and quantum.  
In fact, the translational invariance allows one to separate 
the relative motion of atoms
from the motion of the center of mass.
Contrary to this, the rotational invariance of molecules is not able
to separate the vibrational motion 
from the rotational motion of the whole molecular system,
which was shown by Guichardet\cite{Guichardet}.
The nonseparability of vibration and rotation of a molecule is 
an underlying principle 
that allows cats to fall on their legs when launched in the air.
While they have zero angular momentum and 
is free from external torque during the fall,
they can make a rotation after a vibrational motion.
It was Guichardet\cite{Guichardet} 
who gave a decisive answer to this seemingly strange fact.
He defined a natural connection on the center-of-mass system
and thereby applied the holonomy theorem
to show that the existence of nonvanishing curvature results 
in the nonseparability of vibration and rotation.
It is to be noted here that the center-of-mass system is made into 
a fiber bundle if collinear configurations of the molecule 
are gotten rid of. 
The connection is defined on this restricted center-of-mass system. 

In spite of the nonseparability of vibration and rotation,
the reduction method is still applicable.
One of the authors (T.I.) applied the reduction method of 
Marsden and Weinstein to the molecular system 
by the use of conservation of angular momentum
and gave a reduced Hamiltonian system\cite{Iwai1},
according to which the ``internal'' motion of molecules
is coupled with rotation through a kind of gauge field.
The expression of the reduced equation of motion in terms of 
local coordinates is given in \cite{Iwai4}. 
He also constructed the Schr{\"o}dinger equation
to describe the internal motion of molecules 
in both two and three dimensions,
using differential geometry of fiber bundles 
and connections\cite{Iwai2,Iwai3,Tachibana-Iwai}.
However, in the previous papers\cite{Iwai2,Iwai3,Tachibana-Iwai},
collinear configurations of the molecule
in which all the atoms are aligned in a line in $ \R^3 $
were out of consideration for the reason that
the collinear configurations form a singular point set 
which prevents the center-of-mass system from being made 
into a principal fiber bundle.
The bundle picture is extended to be applicable to 
a system of rigid bodies. 
Classical and quantum mechanics for the system of jointed 
identical axially symmetric cylinders is treated in 
\cite{Iwai5}. 

It is Kummer\cite{Kummer} who first discussed the reduction of 
the cotangent bundle of a principal fiber bundle with a connection. 
He may have been aware of a use of the connection in many-body systems.
However, he did not refer to the center-of-mass system as a 
principal fiber bundle in general, but referred to a planar 
three-body system as an $S^1$-bundle. 
He claimed also that the reduction method had been applied to 
the planar three-body system by Satzer\cite{Satzer}. 
Guichardet is the first who showed that the connection defined 
on the restricted center-of-mass system as a principal fiber 
bundle played an important role in the study of many body systems; 
he showed that the holonomy group of the connection coincides 
with the structure group by the use of the Ambrose-Singer 
holonomy theorem, along with the conclusion that any rotation 
can be realized as a holonomy associated with a closed loop 
in the base manifold (or a shape space). 

This paper has an aim to improve 
the previous theory so that it may be applicable 
on the whole center-of-mass system which includes the singular 
point set stated above. 
Since the center-of-mass system is endowed with a natural 
Riemannian metric and admits the action of $SO(3)$ as an 
isometry group, a general setting to start with is that  
a configuration space $ M $ is a Riemannian manifold on which 
a compact Lie group $ G $ acts by isometry.
{}For a quantum system on $ M $ with symmetry Lie group $ G $,
the reduced quantum system by symmetry is to be defined 
on the quotient space $ Q = M/G $.
However, a problem arises since the $ Q $ is not a smooth manifold 
in general, which may include singular points.
A part of $ Q $, denoted by $ Q_{\mu} $, is a smooth manifold, 
which is called the internal or shape space endowed with 
local coordinates describing the internal degrees of freedom 
of the molecule. 
One can set up a reduced quantum system on the smooth 
manifold $Q_{\mu}$\cite{Iwai2,Iwai3}.  
However, if one considers the whole $ Q $, then
a question arises as to how wave functions of internal coordinates
should behave on the singular point set. 
This article will provide a general formulation to describe
quantum mechanics of a reduced system with singular points 
taken into account. 

Let $ L_2(M) $ be the Hilbert space of square integrable functions 
on the Riemannian manifold $ M $, 
which is to be looked upon as the space of wave functions on the 
center-of-mass system. 
The group action $ G \times M \to M $; $ (g,x) \mapsto gx $
induces a unitary representation of $ G $ in $ L_2(M) $
through the action $ G \times L_2(M) \to L_2(M)$; 
$(g, f(x)) \mapsto f(g^{-1}x) $.
This representation will be decomposed into irreducible ones 
to give rise to representation subspaces of $ L_2(M) $ accordingly.  
To get an idea of the decomposition of $ L_2(M) $, 
the Peter-Weyl theorem on unitary representations of compact groups 
is of great help, 
since the theorem is understood to provide the decomposition 
of the Hilbert spaces of functions on groups. 
The decomposition of $ L_2(M) $ is then performed 
in a similar manner,
to define the spaces of equivariant functions on $ M $  
according to respective irreducible unitary representations of $ G $.
These spaces will give rise to wave functions 
reduced by the use of the angular momentum conservation, 
since choosing an irreducible unitary representation 
of the isometry group amounts 
to keeping an angular momentum eigenvalue fixed.
The equivariant functions will be shown to satisfy 
a good boundary condition, in a natural manner, 
on the set of singular points.
The general formalism for reducing wave functions is applicable 
to molecular mechanics.  
In particular, a triatomic molecule will be studied in detail.

The idea of using the Peter-Weyl theorem for quantization is old;
it dates back to Casimir's work on quantum mechanics of rigid body 
in 1931.
The idea of using the invariant functions to set up a 
reduced quantum system by symmetry is also old; 
for example, the quantum theories of 
electromagnetism\cite{Dirac1} or gravitation\cite{Dirac2,Moncrief}
are described by 
gauge invariant and reparametrization invariant states, respectively.
The idea of using equivariant functions is not new, either;
it has already been introduced by T.I.\cite{Tachibana-Iwai,Iwai3}
in molecular quantum mechanics
and by Landsman and Linden\cite{Landsman-Linden}
in quantization on homogeneous spaces, respectively. 
A point to make in the present paper is to extend the concept
of equivariant functions on a principal fiber bundle 
to any manifold on which a compact group dose not necessarily 
act freely. 

There are other methods of reduction by symmetry.
A more algebraic formulation was set up
by Landsman\cite{Landsman} and Wren\cite{Wren},
who used a representation theory of $ C^* $ algebras 
due to Rieffel\cite{Rieffel}.
The present formulation is rather geometric and comprehensible
in applications.
On the other hand, the path integral formulation is compatible with 
a method of reduction by symmetry.
It has been shown 
\cite{Landsman-Linden,Tanimura1,Tanimura2}
that path integral over a Lie group $ G $
is reduced to path integral over a homogeneous space $ Q = G/H $
when the system possesses symmetry given by a subgroup $ H $.
However, this method is only applicable to homogeneous spaces,
which are free from singularity,
while
the method proposed in the present paper will be applicable
even when the quotient space $ Q = M/G $ is not a smooth manifold, 
as will be shown later.

On the other hand, {\'S}niatycki and Weinstein\cite{Sniatycki},
and other people\cite{Arms} have studied reduction and quantization
of symplectic systems with symmetry.
They consider reduction of symplectic manifold
when the momentum map takes a singular value 
and therefore the level set does not form a smooth manifold.
In our context, singularity refers to
points of a configuration space
which admit larger isotropy groups than those at generic points.
The subject to be considered here is different from what 
has been considered in \cite{Arms,Sniatycki}, 
and a relation between these subjects will not be examined. 

The plan of the present paper is as follows:
In Sec. II, 
the Peter-Weyl theorem on unitary representations of compact groups
is reviewed briefly.
This theorem is extended to be applied to a unitary representation 
of a compact Lie group $ G $ in a Hilbert space $ {\cal H} $ 
in order to decompose $ {\cal H} $ 
into a series of invariant subspaces. 
The decomposition procedure is then applied to
the unitary representation of $ G $ in the Hilbert space $ L_2(M) $.
As far as the decomposition is concerned, 
$ M $ is assumed only to be equipped with a $ G $-invariant measure, 
and does not need to be a Riemannian manifold. 
The above stated decomposition of $ L_2(M) $ proves to be 
a decomposition into a series of spaces 
of ``equivariant" functions. 
Section III  
contains a study of geometric structure of 
the $ G $-manifold $ M $, where the assumption is not yet made 
that $ M $ is a Riemannian manifold. 
It will be shown that the orbit space $ Q = M/G $ becomes
not a smooth manifold but a collection of smooth manifolds 
of various dimensions,
which structure is called stratification.
With this stratification taken into account,
a connection on $ M $, equivariant functions on $ M $,
and covariant derivatives will be defined.
In Sec. IV, 
$ M $ is assumed to be a Riemannian manifold, and 
$ G $ to act on $ M $ by isometry.   
The Laplacian acting on smooth functions on $ M $ will be reduced 
to be defined on each space of equivariant functions according 
to the decomposition of $ L_2(M) $,
which will provide a reduced quantum system on $ Q $.
The reduced systems are well defined. 
In fact, the equivalence condition will provide 
a boundary condition on equivariant functions at singular points. 
As a simple example, the Laplacian on $ \R^2 $ will be studied.
In Sec. V,
the general formalism developed in the preceding sections
is applied to molecular mechanics.
A general setting for $ N $-atomic molecules will be established,
and then 
triatomic molecules will be studied in detail.
Sec. VI contains conclusion and discussion.

\section{Reduction of quantum systems with symmetry}
\subsection{The Peter-Weyl theorem}
We start with a brief review of the Peter-Weyl theorem,
which will provide a key idea to the reduction method 
for quantum dynamical systems with symmetry.

Let $ G $ be a compact Lie group.
Then there is the unique Haar measure $ \mu_G $ on $ G $
which is normalized to satisfy $ \int_G d \mu_G = 1 $.
Let $ L_2 (G) $ denote 
the space of all the square integrable functions on $ G $,
$ L_2 (G) := \{ f : G \to \C \, | \, 
\int_G |f(g)|^2 d \mu_G (g) < \infty\} $,
which is equipped with the inner product
\begin{equation}
	\bra f, f' \ket
	:=
	\int_G \overline{ f(g) } f'(g) d \mu_G (g)
\end{equation}
for $ f $ and $ f' $ of $ L_2 (G) $.

Let $ \rho^\chi : G \to U( {\cal H}^\chi ) $ 
denote an irreducible unitary representation of $ G $ 
on a Hilbert space $ {\cal H}^\chi $
of finite dimension $ d^\chi := \dim {\cal H}^\chi $,
where $ \chi $ is to label all of
inequivalent irreducible unitary representations of $ G $, 
and $ U( {\cal H}^\chi ) $ denotes the space of unitary operators 
on $ {\cal H}^\chi $. 
By $ \rho^\chi_{ij} (g) $ we denote the matrix elements 
of $ \rho^{\chi}(g) $ with respect to some orthonormal basis 
of $ {\cal H}^\chi $,
where indices range over $ i, j = 1, 2, \cdots , d^\chi $.

The Peter-Weyl theorem then states that
all the functions 
$ \{ \sqrt{d^\chi} \, \rho^\chi_{ij} (g) \}_{ \chi,i,j } $ 
form a complete orthonormal set (CONS) in $ L_2(G) $.
Namely, one has the orthonormality relations 
\begin{equation}
	d^\chi
	\int_G 
	\overline{\rho^\chi_{ij}(g)} \rho^{\chi'}_{k \ell}(g) 
	d \mu_G(g)
	=
	\delta^{\chi \chi'} \, \delta_{i k} \, \delta_{j \ell}
	\label{Peter-Weyl1}
\end{equation}
along with the completeness condition that if
\begin{equation}
	\int_G \overline{\rho^\chi_{ij}(g)} f(g) d \mu_G(g) = 0 
	\label{Peter-Weyl2}
\end{equation}
for all $ \chi, i, j $, then $ f \equiv 0 $.
Hence, any function $ f \in L_2(G) $ can be expanded in a 
Fourier series in  
$ \{ \sqrt{d^\chi} \, \rho^\chi_{ij} \}_{ \chi,i,j } $, so that one has 
\begin{eqnarray}
	f(h)
	& = &
	\sum_{\chi, i, j}
	d^\chi
	\rho^\chi_{ij} (h)
	\int_G
	\overline{ \rho^\chi_{ij} (g) }
	f(g) 
	d\mu_G(g)
	\nonumber
	\\
	& = &
	\sum_{\chi, i}
	d^\chi
	\int_G
	\rho^\chi_{ii} (h g^{-1})
	f(g) 
	d\mu_G(g)
	\nonumber
	\\
	& = &
	\sum_{\chi, i}
	d^\chi
	\int_G
	\rho^\chi_{ii} (g)
	f( g^{-1} h ) 
	d\mu_G(g).
	\label{Peter-Weyl3}
\end{eqnarray}
The expansion formula (\ref{Peter-Weyl3}) is put formally in a 
compact form, 
\begin{equation}
	\sum_{\chi, i} d^\chi \, \rho^\chi_{ii} (g) = \delta(g),
	\label{delta on G}
\end{equation}
where $ \delta(g) $ is  Dirac's delta function on $ G $
with respect to the measure $ \mu_G $. 
The Peter-Weyl theorem also implies that 
\begin{equation}
   L_2(G) \cong \bigoplus_{\chi}
   \bigl(({\cal H}^\chi)^*\otimes {\cal H}^\chi\bigr),
   \label{decomp on G}
\end{equation}
where it is to be noted that $ ({\cal H}^\chi)^*\otimes {\cal H}^\chi $ 
is isomorphic with the direct sum of $ d^{\chi}$ copies 
of ${\cal H}^{\chi} $.

\subsection{Method of reduction}
We now apply the Peter-Weyl theorem to 
a quantum dynamical system with symmetry 
to obtain a series of reduced systems. 
A quantum dynamical system is defined to be a pair $({\cal H},H)$ of  
a Hilbert space $ {\cal H}$ and a Hamiltonian $ H $, 
where $H$ is a self-adjoint operator on $ {\cal H} $.
Suppose that the system $ ({\cal H}, H) $ admits
a compact Lie group $ G $ as a symmetry group, 
namely, each element $ g \in G $ is represented as 
a unitary operator $ U(g) $ which acts on $ {\cal H} $,  
commuting with $ H $.

In view of (\ref{Peter-Weyl3}), we define, 
for each label $ (\chi,i) $, an operator $ P^\chi_i $ 
on $ {\cal H} $ to be 
\begin{equation}
	P^\chi_i 
	:=
	d^\chi
	\int_G
	\rho^\chi_{ii}(g) U(g) d \mu_G(g),
	\label{projection}
\end{equation}
which apparently commutes with $ H $.
{}From the $G$-invariance of the measure $ \mu_G $ and from 
the orthonormality relations (\ref{Peter-Weyl1}), it follows that 
\begin{equation}
	( P^\chi_i )^\dagger = P^\chi_i,
	\qquad
	P^\chi_i P^{\chi'}_j 
	= \delta^{\chi \chi'} \delta_{ij} \, P^\chi_i. 
	\label{orthonormality}
\end{equation}
Further, the completeness condition (\ref{delta on G}) implies that 
\begin{equation}
	\sum_{\chi, i} P^\chi_i = \mbox{id}_{\cal H},
	\label{completeness}
\end{equation}
where $\mbox{id}_{\cal H}$ is the identity operator on ${\cal H}$. 
Equations (\ref{orthonormality}) and (\ref{completeness}) show  
that the set $ \{ P^\chi_i \}_{\chi, i} $ forms a family of 
orthogonal projection operators on $ {\cal H} $, bringing about 
the orthogonal decomposition of $ {\cal H} $,
\begin{equation}
	{\cal H} =
	\bigoplus_{\chi, i} \, \Image \, P^\chi_i,
	\label{decomposition}
\end{equation}
which is an analog to (\ref{decomp on G}).  
Moreover,  each subspace $ \Image \, P^\chi_i $ is invariant 
under the action of the Hamiltonian $ H $. 
Thus the dynamical system $ ({\cal H},H) $ is broken up into 
a family of subsystems $ ( \Image \, P^\chi_i, H ) $  
labeled by $ (\chi, i) $. 
We call each system $ ( \Image \, P^\chi_i, H ) $ 
a reduced quantum dynamical system.

In the language of physics,
the pair $ ( \chi, i ) $ labels conserved quantities associated 
with the symmetry group $ G $, and thereby define a closed dynamical 
system that consists of the states with the assigned conserved 
quantities. 
{}For example, if the original system has $ G = SU(2) $ symmetry,
the angular momentum is conserved. 
The states labeled by $ (j,m) $ have the total angular momentum and 
the component of the angular momentum fixed at $ J^2=j(j+1) $ and  
$ J_3=m $, respectively, and are described as vector-valued functions  
with $2j+1$ components.  

\subsection{Characterization of the reduced system}
To gain a deeper insight into the decomposition (\ref{decomposition}),  
we introduce an operator on $ {\cal H} $ by  
\begin{equation}
	V^\chi_{ij}
	:=
	d^\chi
	\int_G
	\rho^\chi_{ij}(g) U(g) d \mu_G(g)
	\label{V_ij}
\end{equation}
for each label $ \chi $ and indices $ i,j = 1, \cdots, d^\chi $.
In particular, one has $ V^\chi_{ii} = P^\chi_i $.
A straightforward calculation shows that 
\begin{equation}
	( V^\chi_{ij} )^\dagger = V^\chi_{ji},
	\qquad
	V^\chi_{ij} \, V^{\chi'}_{k \ell} 
	= \delta^{\chi \chi'} \delta_{jk} \, V^\chi_{i \ell}.
	\label{orthonormality of V}
\end{equation}
As an immediate consequence, one obtains $ V_{ij}^{\chi}P_k^{\chi}=0 $  
if $ k\neq j $, so that the domain of $ V_{ij}^{\chi} $ reduces 
naturally to $ \Image P_j^{\chi} $ on account of (\ref{decomposition}). 
Since
$ ( V^\chi_{ij} )^\dagger V^\chi_{ij} 
=   V^\chi_{ji}           V^\chi_{ij} 
=   V^\chi_{jj}
=   P^\chi_j $,
it holds that $ \Kernel V^\chi_{ij} = \Kernel P^\chi_j $.
Similarly, from 
$   V^\chi_{ij} ( V^\chi_{ij} )^\dagger 
=   V^\chi_{ij}   V^\chi_{ji} 
=   V^\chi_{ii}
=   P^\chi_i $,
it follows that $ \Kernel (V^\chi_{ij})^{\dagger} = \Kernel P^\chi_i $,  
and thereby that  
$ \Image \, V^\chi_{ij} 
= ( \Kernel (V^\chi_{ij})^\dagger )^\perp
= ( \Kernel \, P^\chi_i )^\perp
=   \Image  (P^\chi_i)^\dagger 
=   \Image \, P^\chi_i $.
Therefore, $ V^\chi_{ij} $ becomes a unitary transformation
\begin{equation}
	V^\chi_{ij} : \Image \, P^\chi_j \to \Image \, P^\chi_i.
	\label{unitarity of V}
\end{equation}

As a collection of $ V^{\chi}_{ij}$, we define another operator
$ V^\chi : 
{\cal H}^\chi \otimes {\cal H} \to {\cal H}^\chi \otimes {\cal H} $ by
\begin{equation}
	V^\chi
	:=
	d^\chi
	\int_G
	\rho^\chi(g) \otimes U(g) d \mu_G(g). 
	\label{V}
\end{equation}
Then Eq.(\ref{orthonormality of V}) implies that 
\begin{equation}
	( V^\chi )^\dagger = V^\chi,
	\qquad
	V^\chi \, V^\chi = d^\chi \, V^\chi,
	\label{VV}
\end{equation}
which shows that $ V^\chi/d^\chi $ is 
a projection operator on $ {\cal H}^\chi \otimes {\cal H} $.
Further, it is easy to verify that 
$ V^\chi $ satisfies, for any $ h \in G $, 
\begin{eqnarray}
	V^\chi ( \rho^{\chi}(h) \otimes U(h) ) & = & V^\chi,
	\label{R-invariance of V} \\
	( \rho^{\chi}(h) \otimes U(h) ) V^\chi & = & V^\chi,
	\label{L-invariance of V}
\end{eqnarray}
which implies that  
\begin{equation}
	 \Image \, V^\chi 
	=
	( {\cal H}^\chi \otimes {\cal H} )^G
	:=
	\{ \psi \in {\cal H}^\chi \otimes {\cal H} 
	\, | \,
	( \rho^{\chi}(h) \otimes U(h) ) \psi = \psi,
	\, \forall h \in G \}. 
	\label{Image of V}
\end{equation}
We call $ ( {\cal H}^\chi \otimes {\cal H} )^G $ the subspace 
of invariant vectors of $ {\cal H}^\chi \otimes {\cal H} $.

Let $ \{ e^\chi_1, \cdots, e^\chi_{d^\chi} \} $ 
be an orthonormal basis of $ {\cal H}^\chi $,
which defines an injection $ e^\chi_i : \C \to {\cal H}^\chi $
by $ \lambda \mapsto \lambda e^\chi_i $ 
for each $ i =1, \cdots, d^\chi $.
Its adjoint operator
$ (e^\chi_i)^\dagger : {\cal H}^\chi \to \C $
is defined by a surjection
$ v \mapsto \bra e^\chi_i, v \ket ,\;v\in {\cal H}^{\chi}$. 
The domain of respective maps extends to the tensor product 
space with $ {\cal H} $ to give rise to 
$ e^\chi_i : {\cal H} \to {\cal H}^\chi \otimes {\cal H} $
and
$ (e^\chi_i)^\dagger : {\cal H}^\chi \otimes {\cal H} \to {\cal H} $.
With these notations, $ V^{\chi}_{ij} $ is put in the form  
$ V^\chi_{ij} = (e^\chi_i)^\dagger \, V^\chi \, e^\chi_j $.
We then introduce an operator by  
\begin{equation}
	S^\chi_j 
	:= \dfrac{1}{\sqrt{d^\chi}} \, V^\chi \, e^\chi_j :
	{\cal H} \to {\cal H}^\chi \otimes {\cal H}.
	\label{I}
\end{equation}
The adjoint operator is expressed as 
\begin{equation}
	(S^\chi_j)^\dagger
	:= \dfrac{1}{\sqrt{d^\chi}} \, (e^\chi_j)^\dagger \, V^\chi :
	{\cal H}^\chi \otimes {\cal H} \to {\cal H}.
	\label{I^dagger}
\end{equation}
Then the second relation of (\ref{VV}) yields 
\begin{equation}
	(S^\chi_j)^\dagger S^\chi_j
	= 
	(e^\chi_j)^\dagger \, V^\chi \, e^\chi_j
	=
	P^\chi_j.
	\label{I^dagger I}
\end{equation}
On the other hand, from (\ref{orthonormality of V}),   
we observe that 
$ V^\chi \, e^\chi_j \, (e^\chi_j)^\dagger \, V^\chi = V^{\chi} $, 
and thereby obtain   
\begin{equation}
	S^\chi_j (S^\chi_j)^\dagger
	= 
	\dfrac{1}{d^\chi} \, 
	V^\chi \, e^\chi_j \, (e^\chi_j)^\dagger \, V^\chi 
	=
	\dfrac{1}{d^\chi} \, 
	V^\chi.
	\label{I I^dagger}
\end{equation}
Since $ P^{\chi}_j $ and $ V^{\chi}/d^{\chi} $ are projection 
operators on $ {\cal H} $ and on $ {\cal H}^{\chi}\otimes {\cal H} $, 
respectively, Eqs.(\ref{I^dagger I}) and (\ref{I I^dagger}) are 
put together to imply that 
the restricted map
\begin{equation}
	S^\chi_j : 
	\Image \, P^\chi_j 
	\isoarrow
	\Image \, V^\chi = ({\cal H}^\chi \otimes {\cal H})^G
	\label{reduced subspace}
\end{equation}
is a unitary transformation. 
Thus the subspace $ \Image \, P^\chi_j $ is characterized as  
$ ({\cal H}^\chi \otimes {\cal H})^G $.

We turn to the Hamiltonian $ H $ acting on $ {\cal H} $. 
It appears that $ H $ is extended to an operator 
$ \mbox{id}_{{\cal H}^{\chi}} \otimes H $ 
on $ {\cal H}^\chi \otimes {\cal H} $.
Since $ H $ commutes with $ U(g) $ for each $ g \in G $,
the extended operator $ \mbox{id}_{{\cal H}^{\chi}} \otimes H $ 
commutes with $ V^\chi $ and with $ S^\chi_j $ as well, so that 
one has 
\begin{equation}
	S^\chi_j H = ( \mbox{id}_{{\cal H}^{\chi}} \otimes H ) S^\chi_j.
	\label{commutativity}
\end{equation}
It then turns out that the reduced system $ ( \Image \, P^\chi_j, H ) $ 
is identified with
$ ( ({\cal H}^\chi \otimes {\cal H})^G, \mbox{id}_{{\cal H}^{\chi}}
 \otimes H ) $,
which will be investigated in the following sections in detail.

Before investigation into the reduced system, we wish to consider 
how the action of $ G $ is decomposed according to  
the decomposition of $ {\cal H} $.   
In the below, the dual space $ ({\cal H}^\chi)^* $ to 
$ {\cal H}^\chi $ is taken as 
$ ({\cal H}^\chi)^* := \mbox{Hom} \, ({\cal H}^\chi; \C) $,  
irrespective of the inner product in $ {\cal H}^{\chi} $. 
Then the map $ \sqrt{d^\chi} \, (S^\chi_j)^\dagger 
= (e^\chi_j)^\dagger \, V^\chi :
{\cal H}^\chi \otimes {\cal H} \to {\cal H} $
can be viewed as a map 
$ T^\chi_j : {\cal H} \to ({\cal H}^\chi)^* \otimes {\cal H} $ 
in the manner as follows: 
Using the identification 
$ \mbox{End} \, {\cal H}^\chi \cong 
({\cal H}^\chi)^* \otimes {\cal H}^\chi $,
we may regard the map
$ V^\chi : 
{\cal H}^\chi \otimes {\cal H} \to {\cal H}^\chi \otimes {\cal H} $
as a map
$ V^\chi : 
{\cal H} \to 
({\cal H}^\chi)^* \otimes {\cal H}^\chi \otimes {\cal H} $.
Combining $ \mbox{id}_{({\cal H}^{\chi})^*}\otimes(e^\chi_j)^\dagger : 
({\cal H}^{\chi})^* \otimes {\cal H}^\chi \otimes {\cal H} 
\to ({\cal H}^{\chi})^* \otimes {\cal H} $ with $ V^\chi $, 
we can express $ T^{\chi}_j $ as the map 
\begin{equation}
	T^\chi_j 
	= (\mbox{id}_{({\cal H}^{\chi})^*}\otimes (e^\chi_j)^\dagger) V^\chi : 
	{\cal H} \to ({\cal H}^\chi)^* \otimes {\cal H}.
	\label{T}
\end{equation}
Then it can be verified that
\begin{equation}
	(T^\chi_j)^\dagger T^\chi_j
	=
	\sum_k P^\chi_k,
	\qquad
	T^\chi_j (T^\chi_j)^\dagger
	= 
	\mbox{id}_{({\cal H}^{\chi})^*} \otimes P^\chi_j, 
	\label{TT}
\end{equation}
where $ (T^{\chi}_j)^\dagger: ({\cal H}^{\chi})^*\otimes {\cal H} 
\to {\cal H} $ is the adjoint operator.  
{}From this, it follows that $ T^\chi_j $ yields a unitary transformation
\begin{equation}
	T^\chi_j : 
	\bigoplus_k \Image \, P^\chi_k
	\isoarrow
	({\cal H^\chi})^* \otimes \Image \, P^\chi_j.
	\label{unitarity of T}
\end{equation}
Then the right invariance of $ V^\chi $ under the $ G $ action, 
expressed in (\ref{R-invariance of V}),  
makes the following diagram commutative,  
\begin{equation}
	\begin{array}{ccccc}
	  {\cal H}
	& =
	& \bigoplus_{\chi, i} \, \Image \, P^\chi_i
	& \mapright{ T^\chi_j }{}
	& ({\cal H}^\chi)^* \otimes \Image \, P^\chi_j
	\\
	  \mapdown{ U(h) }{}
	& & & 
	& \mapdown{}{ %
	{}^t {\rho^\chi} (h^{-1}) \otimes \mbox{\scriptsize id} }
	\\
	  {\cal H}
	& =
	& \bigoplus_{\chi, i} \, \Image \, P^\chi_i
	& \mapright{ T^\chi_j }{}
	& ({\cal H}^\chi)^* \otimes \Image \, P^\chi_j,
	\end{array}
	\label{diagram of representation}
\end{equation}
where 
$ {}^t {\rho^\chi} (h^{-1}) = \overline{ {\rho^\chi} (h) } $ 
is the contragredient representation of $ G $ 
on $ ({\cal H}^\chi)^* $.
As for the action of the Hamiltonian $ H $, we obtain the 
following commutative diagram accordingly, 
\begin{equation}
	\begin{array}{ccccc}
	  {\cal H}
	& =
	& \bigoplus_{\chi, i} \, \Image \, P^\chi_i
	& \mapright{ T^\chi_j }{}
	& ({\cal H}^\chi)^* \otimes \Image \, P^\chi_j
	\\
	  \mapdown{ H }{}
	& & & 
	& \mapdown{}{ %
	  \mbox{\scriptsize id} \otimes H }
	\\
	  {\cal H}
	& =
	& \bigoplus_{\chi, i} \, \Image \, P^\chi_i
	& \mapright{ T^\chi_j }{}
	& ({\cal H}^\chi)^* \otimes \Image \, P^\chi_j .
	\end{array}
	\label{diagram of H}
\end{equation}
The commutative diagrams (\ref{diagram of representation}) 
and (\ref{diagram of H}) show that 
the decomposition of the representation $ ( G, {\cal H}, U ) $
is compatible with the spectral resolution of $ H $.

\subsection{Equivariant functions}
The general reduction method introduced in the previous section 
applies to a quantum system on a configuration space $ M $
which admits the action of a compact Lie group $ G $.

Suppose that a compact Lie group $ G $ 
acts on a differentiable manifold $ M $ by diffeomorphisms,
namely, we are given a $ C^\infty $ map $ G \times M \to M $
satisfying $ e x = x $ and $ ( g h ) x = g ( h x ) $ for
any $ x \in M $ and any $ g, h \in G $ 
with the identity element $ e \in G $.
Then $ M $ is called a {\it G-manifold}.
In addition, suppose that $ M $ is equipped with
a measure $ \mu_M $ which is invariant under the action of $ G $.
Let $ {\cal H} = L_2(M) $
be the space of square integrable functions on $ M $
with respect to the measure $ \mu_M $.
The group $ G $ is represented on $ {\cal H} = L_2(M) $ 
by unitary operators $ U(h),\;h\in G $, through
\begin{equation}
	( U(h) f )(x) := f( h^{-1} x ),
	\qquad
	f \in L_2(M).
	\label{U}
\end{equation}
Then the reduction method applies to $ {\cal H}=L_2(M) $ 
to yield the Hilbert subspace $ \Image P^\chi_j $.  
The unitary transformation $ S^\chi_j $  
given in (\ref{reduced subspace}) allows us to identify  
$ \Image \, P^\chi_j \: ( \subset L_2(M) ) $ 
with
$ ( {\cal H}^\chi \otimes L_2(M) )^G \:
( \subset {\cal H}^\chi \otimes L_2(M) ) $.
The space $ {\cal H}^\chi \otimes L_2(M) $ can be identified with 
the Hilbert space, $ L_2(M;{\cal H}^\chi) $, of square integrable 
$ {\cal H}^{\chi} $-valued functions on $ M $; 
\begin{equation}
	L_2(M;{\cal H}^\chi)
	:=
	\{ \psi : M \to {\cal H}^\chi
	\, | \,
	\int_M || \psi(x) ||^2 \, d \mu_M (x) < \infty \}, 
	\label{L_2(M;H)}
\end{equation}
which is equipped with the inner product
\begin{equation}
	\bra \phi, \psi \ket 
	:=
	\int_M \bra \phi(x), \psi(x) \ket d \mu_M (x),
	\qquad
	\phi, \psi \in L_2(M; {\cal H}^\chi),
	\label{inner product}
\end{equation}
where $ \bra \phi(x), \psi(x) \ket $ denotes the inner product 
in $ {\cal H}^{\chi} $. 
Then, the condition given in (\ref{Image of V}) 
along with $ (\ref{U}) $ implies that 
$ \psi \in ( {\cal H}^\chi \otimes L_2(M) )^G $, when viewed 
as an $ {\cal H}^{\chi} $-valued function, satisfies
\begin{equation}
	 \psi (hx) 
	= \rho^\chi(h) \psi (x), \quad h\in G,
	\label{equivariance of psi}
\end{equation}
which shows that $ \psi $ is equivariant under the $ G $-action. 
We conclude therefore that the reduced Hilbert space $ \Image P^\chi_j $ 
is identified with the space of square integrable equivariant functions, which we denote by $ L_2(M;{\cal H}^{\chi})^G $;
\begin{equation}
\begin{array}{lll}
 ( {\cal H}^\chi \otimes L_2(M) )^G & \cong & 
L_2(M;{\cal H}^{\chi})^G \\[2mm]
 & = & 
\{\psi \in L_2(M;{\cal H}^{\chi})|\;
	 \psi (gx) 
	= \rho^\chi(g) \psi (x), \quad g\in G \}. 
\end{array}
        \label{equiv function}
\end{equation}
Here we have to note that according to the decomposition 
(\ref{decomposition}) along with (\ref{reduced subspace}), 
(\ref{unitarity of T}), and (\ref{equiv function}), 
the Hilbert space $ L_2(M) $ is decomposed into 
\begin{equation}
 L_2(M)\cong  \bigoplus_{\chi}
 \bigl(({\cal H}^{\chi})^*\otimes L_2(M;{\cal H}^{\chi})^G\bigr).
 \label{decomp on M}
\end{equation}

So far we have characterized the reduced Hilbert space 
on the $ G $-manifold $ M $. 
We now have to specify the Hamiltonian on $ L_2(M) $ and 
to reduce it. 
We will take a Hamiltonian as defined to be the sum
of the Laplacian on $ M $ and a potential energy function on $ M $.
In the succeeding sections,  
we will study the geometric structure of $ M $ 
in order to analyze the Laplacian on $ M $ according to  
the decomposition (\ref{decomp on M}). 

\section{Geometric setting on G-manifolds}

\subsection{Stratification of G-manifolds}
According to Davis\cite{Davis paper}, 
$ G $-manifolds can be viewed as collections of fiber bundles.
We here make a brief review of his idea in a suitable form
for our application.
{}For more rigorous definitions, see the 
literature\cite{Davis lecture}.

Let $ M $ be a $ G $-manifold. 
For a point $ x \in M $, we denote 
the {\it isotropy subgroup} at $ x $ and 
the {\it G-orbit} of $ x $ by 
$ G_x := \{ g \in G \, | \, gx = x \} $ and 
by $ {\cal O}_x := \{ gx \, | \, g \in G \} $, 
respectively.  
Then one has $ {\cal O}_x \cong G/G_x $.

Take $ G_x $ and $ G_y $ for two points $ x, y \in M $.
If $ G_x $ is conjugate to $ G_y $ 
by an inner automorphism
$ A_g : G \to G $; $ h \mapsto g h g^{-1} $
with some $ g \in G $; 
$ G_y = g \cdot G_x \cdot g^{-1} $,
then the orbits $ {\cal O}_x $ and $ {\cal O}_y $ 
are diffeomorphic to each other
by the correspondence induced by $ A_g $.
Of course, if $ y $ is in the orbit $ {\cal O}_x $ , 
namely, if there exists some $ g \in G $ such that $ y = gx $,
then $ G_y $ is conjugate to $ G_x $ by $ A_g $.
The conjugacy class of $ G_x $ 
is called the orbit type of $ {\cal O}_x $
and denoted by $ [G_x] $ or $ \tau $.
We say that the point $ x $ itself also has the orbit type $ \tau $, 
if $ {\cal O}_x $ has the orbit type $ \tau $.
Let $ {\cal T} (M) $ denote the set of all the orbit types appearing
in $ M $; 
$ {\cal T}(M) := \{ [G_x] \, | \, x \in M \} $.
{}For each $ \tau \in {\cal T}(M) $,
we denote a representative of the conjugacy class by 
$ G_\tau \in \tau $.

One can introduce a partial order on $ {\cal T}(M) $ as follows:
We say that 
$ \tau_1 $ is lower than $ \tau_2 $ 
$ ( \tau_1, \tau_2 \in {\cal T}(M) ) $,
$ \tau_1 \le \tau_2 $,
if there are representatives 
$ G_{\tau_1} \in \tau_1 $ and
$ G_{\tau_2} \in \tau_2 $ such that
$ G_{\tau_1} \supset G_{\tau_2} $.
An orbit $ {\cal O}_x $ is called {\it maximal}
if its orbit type $ \tau $ is maximal with respect to this order.
Orbits which are not maximal are called {\it singular}.
We say also that a point $ x $ is maximal or singular
according to whether the orbit type of $ {\cal O}_x $ 
is maximal or singular.
{}For each orbit type $ \tau \in {\cal T}(M) $,
we denote by $ M_\tau := \{ x \in M \, | \, G_x \in \tau \} $
the set of points with the same orbit type $ \tau $, 
which becomes a smooth manifold.
Thus $ M $ is {\it stratified} into a collection of the smooth 
submanifolds $ M_\tau $,
$ M = \coprod_\tau M_\tau $,
which is partially ordered by the system of orbit types 
$ ( {\cal T}(M), \le ) $.
Each manifold $ M_\tau $ is called a {\it stratum}.

The set of orbits $ Q := M/G $ becomes a topological space
with respect to the quotient topology
which is defined by demanding that 
the canonical projection map $ \pi : M \to M/G $ is continuous.
The space $ Q $ is called an {\it orbit space}.
A point $ q \in Q $ is also classified by the orbit type of 
$ \pi^{-1} (q) $.
The point $ q $ is said to be maximal or singular
according to whether the orbit is maximal or singular.
The $ Q $ inherits differentiable structure from $ M $, 
if and only if all the orbits have the same orbit type. 
If otherwise, $ Q $ is stratified 
into a collection of smooth manifolds of various dimensions.
Setting $ Q_\tau := \pi(M_\tau) $, one has $ Q = \coprod_\tau Q_\tau $.
The restriction of $ \pi : M \to Q $ onto each stratum 
$ \pi_\tau : M_\tau \to Q_\tau $ defines a fiber bundle 
with fiber $ G/G_\tau $.
We then call the pentaplet $ (M, G, Q, \pi, {\cal T}(M)) $ 
a {\it stratified bundle}.

If every point of $ M $ has the same orbit type, 
$ \pi : M \to Q $ is nothing but a usual fiber bundle.
In particular, if the group $ G $ acts on $ M $ freely,
namely, $ G_x = \{ e \} $ for all $ x \in M $,
then $ \pi : M \to Q $ becomes a principal fiber bundle
with structure group $ G $.
In this sense, we may regard the $ G $-manifold
as a generalization of fiber bundles
although the base space $ Q=M/G $ is not a smooth manifold.
Mechanics of molecules provides an example of
a stratified bundle that is not a fiber bundle,
as shown in the later section.

According to the principal orbit theorem\cite{Bredon,Davis paper}, 
if the orbit space $ Q = M/G $ is connected, 
there is the {\it maximum} 
orbit type in $ {\cal T}(M) $ with respect to the order $ \le $.
Although the maximum orbit type is also named 
the principal orbit type in \cite{Bredon,Davis paper}, 
we will call it maximum in this paper. 
We assume 
that $ Q $ is connected and that the maximum orbit exists.
We denote the maximum orbit type by $ \mu $.
Moreover, the principal orbit theorem\cite{Bredon,Davis paper} 
states that the maximum stratum
$ M_\mu $ is open and dense in $ M $.
Thus the set $ M_{\mbox{\scriptsize sing}} := M - M_\mu $
of all the singular points coincides with the boundary 
$ \partial M_\mu $.
The image of $ M_\mu $ and $ \partial M_\mu $
by the projection $ \pi $ are denoted by
$ Q_\mu := \pi(M_\mu) $ and
$ \partial Q_\mu := \pi(\partial M_\mu) $, respectively.
We put $ \dim M = m $ and $ \dim Q_\mu = n = m - p $.
Then the dimension of the maximum orbit is $ \dim G/G_\mu = p $.

Though the orbit space $ Q $ is not a manifold,
one can speak of differentiability of functions on $ Q $.
A function of $ \varphi : Q \to \R $ is called of $ C^r $ class
when $ \varphi \circ \pi : M \to \R $ 
is a differentiable function of $ C^r $ class.
Clearly, $ \varphi \circ \pi $ is a $ G $-invariant function on $ M $.
Conversely, any $ G $-invariant function $ f $ on $ M $
is identified with a function on $ Q $.
We denote 
the space of $ G $-invariant functions on $ M $ of $ C^r $ class by
$ C^r (M)^G = 
\{ f : M \to \R \, | \, f(gx) = f(x), \,
\forall g \in G, \, \forall x \in M
\} $.

A tangent vector $ X $ to $ M $ at $ x $ is usually defined 
as a differential operator acting on $ C^\infty(M) $;
$ X : C^\infty(M) \to \R $.
A tangent vector on $ Q $ is defined as follows: 
We define an equivalence relation $ \sim $ in 
the tangent vector space $ T_xM $ by stating 
that $ X \sim Y $ if $ Xf = Yf $ for all $ f \in C^\infty(M)^G $.
The equivalence class of $ X $ is denoted by $ \pi_*(X) $,
which defines a linear map $ \pi_*(X) : C^\infty(M)^G \to \R $.
The set $ T_q Q := \{ \pi_*(X) \, | \, X \in T_x M, \: \pi(x) = q \} $
becomes a vector space 
through the structure of the vector space $ T_x M $
and is called a tangent vector space at $ q \in Q $.

\subsection{Stratified connection}
Let $ (M, G, Q, \pi, {\cal T}(M)) $ be
the stratified bundle defined above.
If $ G $ acts on $ M $ freely,
the stratified bundle becomes a principal fiber bundle.
Although the concept of connection is usually defined
on principal fiber bundles,
we would here like to define extended connections
on stratified bundles.

Let us call a subspace 
$ V_x := 
T_x {\cal O}_x $ 
of $ T_x M $
a {\it vertical subspace} at $ x \in M $.
The action of $ g \in G : M \to M $; $ x \mapsto gx $
induces an action $ g_* : T_x M \to T_{gx} M $ by differentiation.
The complement $ H_x $ of $ V_x $
is called a {\it horizontal subspace}. 
If the direct sum decomposition 
$ T_x M = V_x \oplus H_x $ is smooth
and if the family $ \{ H_x \}_{x \in M} $ satisfies 
the invariance; $ H_{gx} = g_* H_x $, 
the decomposition 
$ T_x M = V_x \oplus H_x $ 
defines a {\it connection}.
However, we should note that
$ \dim V_x = \dim {\cal O}_x $ and 
$ \dim H_x = \dim M - \dim {\cal O}_x $ jump suddenly 
when the point $ x $ passes singular points.
Thus the smoothness of the decomposition
$ T_x M = V_x \oplus H_x $ must be required 
on each stratum $ M_\tau $, so that one understands 
that for any smooth vector field $ X $ which is decomposed
into $ X(x) = X^V(x) + X^H(x) $ according to 
$ T_x M = V_x \oplus H_x $, 
the components $ X^V $ and $ X^H $ are also smooth
on each $ M_\tau $.
Then the assignment $ x \mapsto H_x  $ for each $ x \in M $ is called 
a {\it stratified connection} 
over the stratified bundle $ \pi : M \to Q $.

The decomposition
$ T_x M = V_x \oplus H_x $
induces a decomposition of the dual space
$ T_x^* M = V_x^* \oplus H_x^* $ with
$ V_x^* := 
\{ \phi \in T_x^* M \, | \, 
\phi(v)=0, \, \forall v \in H_x
\} $
and 
$ H_x^* \:= 
\{ \psi \in T_x^* M \, | \, 
\psi(u)=0, \, \forall u \in V_x
\} $,

Let $ \g $ and $ \g_x $ denote 
the Lie algebras of $ G $ and $ G_x $, respectively.
The relation $ \g_{gx} = \Ad_g \g_x $ 
is an immediate consequence of
$ G_{gx} = A_g G_x $.
The group action $ G \times M \to M $; $ (g, x) \mapsto gx $
gives rise to vector fields
$ \g \times M \to TM $; $ ( \xi, x ) \mapsto \xi_M (x) $ 
as infinitesimal transformations.
{}Fixing a point $ x \in M $, one obtains a linear map
$ \theta_x : \g \to T_x M ; \, \theta_x(\xi)=\xi_M(x) $.
It then follows that
$ \Kernel \theta_x = \g_x $ and $ \Image \, \theta_x = V_x $,
and hence that 
$ \theta_x : \g \to T_x M $ induces 
an isomorphism $ \widetilde{\theta}_x : \g / \g_x \isoarrow V_x $.

The connection defined above is described
in term of differential forms.
A connection form $ \omega $ is defined as the composition of
the projection $ T_x M = V_x \oplus H_x \to V_x $
and the inverse map
$ (\widetilde{\theta}_x)^{-1} : V_x \isoarrow \g / \g_x $:
\begin{equation}
	\omega_x :
	T_x M = V_x \oplus H_x \to V_x \isoarrow \g / \g_x
	\label{connection form}
\end{equation}
at each point $ x \in M $.
The form $ \omega $ is thus a one-form which takes values in
quotient spaces of the Lie algebra $ \g $.
It has the following properties:
\begin{eqnarray}
	&&
	\omega( \theta_x (\xi) ) \equiv \xi 
	\quad (\mbox{mod} \, \g_x), 
	\qquad 
	\xi \in \g,
	\label{property1}
	\\
	&&
	( g^* \omega )_x = \Ad_g \omega_x, \qquad 
	g \in G,
	\label{property2}
\end{eqnarray}
where 
$ g^* $ is the pull-back associated with
the map $ g_* : T_x M \to T_{gx} M $
and $ \Ad_g $ is to be understood as a map
$ \Ad_g : \g / \g_x \to \g / \g_{gx} $.
To verify (3.3), we need the formula that 
$\theta_{gx}(\Ad_g \xi)=g_*(\theta_x(\xi))$ for $\xi \in \g$ and 
that $(\widetilde{\theta}_{gx})^{-1}\circ g_* = 
\Ad_g (\widetilde{\theta}_x)^{-1}$. 
The properties (\ref{property1}) and (\ref{property2}) 
are generalization of the well-known defining 
properties of usual connection forms.
It is also noted that the composition map
\begin{equation}
	\widetilde{\theta_x} \circ \omega_x : 
	T_x M = V_x \oplus H_x \to V_x 
	\label{property3}
\end{equation}
is a projection, and that Eq.(3.2) is equivalently written as 
$ \omega_x \circ \widetilde{\theta}_x = 
\mbox{id}_{\g/\g_x}$. 

\subsection{Equivariant forms}
Let $ \psi \in C^\infty(M, {\cal H}^\chi)^G $
be a smooth equivariant function, that is, one has 
$ \psi(gx)=\rho^{\chi}(g)\psi(x)$ for $ g \in G $ (see (\ref{equivariance of psi})).  
If a point $ x \in M $ carries 
a nontrivial isotropy group $ G_x $, then 
the value of $ \psi(x) $ becomes invariant under 
the action of $ G_x $; 
$ \psi(x) = \rho^\chi(g) \psi(x) $ for $ g \in G_x $.
For a subgroup $ G_1 \subset G $, 
we here define $ ({\cal H}^\chi)^{G_1} $
to be a {\it maximum subspace of invariant vectors}
under the action of $ G_1 $, that is,
$ ({\cal H}^\chi)^{G_1} := \{ v \in {\cal H}^\chi \, | \,
\rho^\chi(g) v = v, \, \forall g \in G_1 \} $.
With this notation, we then have  
$ \psi(x) \in ({\cal H}^\chi)^{G_x} $.

Properties of subspaces of invariant vectors are worth remarking.
One has obviously $ ({\cal H}^\chi)^{ \{ e \} }  = {\cal H}^\chi $.
Further, one obtains $ ({\cal H}^\chi)^{G} = \{ 0 \} $, 
if and only if there is no nontrivial invariant vector. 
To a sequence of subgroups
$ \{ e \} \subset G_1 \subset G_2 ( \subset G ) $,
there corresponds a sequence of subspaces
$ {\cal H}^\chi 
\supset ({\cal H}^\chi)^{G_1} 
\supset ({\cal H}^\chi)^{G_2}
( \supset \{ 0 \} ) $.
{}For conjugate subgroups $ G_1 $ and $ A_g G_1 $,
it holds that 
$ ({\cal H}^\chi)^{A_g G_1} = \rho^\chi(g) ({\cal H}^\chi)^{G_1} $.
In particular, on an orbit $ {\cal O}_x $, one has 
$ ({\cal H}^\chi)^{G_{gx}} = \rho^\chi(g) ({\cal H}^\chi)^{G_{x}} $
on account of $ G_{gx} = A_g G_x $.

Like equivariant functions, we can define 
equivariant differential forms in a similar manner.
An {\it equivariant k-form} is defined as 
an $ {\cal H}^\chi $-valued differential $ k $-form
$ \alpha : \wedge^k TM \to {\cal H}^\chi $ satisfying
$ g^* \alpha = \rho^\chi(g) \alpha $ for any $ g \in G $.
When an equivariant $k$-form $ \alpha $ further satisfies
$ i( \xi_M ) \alpha = 0 $ for any $ \xi \in \g $,
it is called an {\it equivariant horizontal \mbox{$k$}-form}.
While the set of all the 
differential $ k $-form on $ M $
is denoted by $ {\mit\Omega}^k (M) $,
the set of all the $ {\cal H}^\chi $-valued differential $ k $-forms
and the set of all the equivariant horizontal $ k $-forms 
are denoted by 
$ {\cal H}^\chi \otimes {\mit\Omega}^k (M) 
\cong {\mit\Omega}^k (M; {\cal H}^\chi) $
and $ {\mit\Omega}_H^k (M; {\cal H}^\chi)^G $, respectively.

At each point $ x \in M $ 
carrying a nontrivial isotropy group $ G_x $,
the equivariant horizontal $k$-form $ \alpha $ takes  
a restricted range as well as equivalent functions. 
It turns out that 
$ \alpha_x ( X_1, \cdots, X_k ) \in ({\cal H}^\chi)^{G_x} $
for any $ X_1, \cdots, X_k \in T_x M $ when $ G_x $ is connected.
The proof is given below:
Let $ L_X $ denote the Lie derivation 
by a vector field $ X $ on $ M $,
and $ \rho^\chi_* $ denote a representation of 
the Lie algebra $ \g $ on $ {\cal H}^\chi $
induced by differentiation
of the representation $ \rho^\chi $ of $ G $.
The defining property of the equivariant form
$ g^* \alpha = \rho^\chi(g) \alpha $ for $ g \in G $ 
is differentiated to give
$ L_{\xi_M} \alpha = \rho^\chi_* (\xi) \alpha $ for $ \xi \in \g $.
This equation and the horizontal condition 
$ i(\xi_M) \alpha = 0 $ for $ \xi \in \g $
are put together with the Cartan formula
$ L_{\xi_M} \alpha = i(\xi_M) d \alpha + d i(\xi_M) \alpha $
to provide
\begin{equation}
	\rho^\chi_* (\xi) \alpha = i(\xi_M) d \alpha,
	\qquad
	\xi \in \g.
	\label{Cartan}
\end{equation}
If $ \xi \in \g_x $, 
one has $ \xi_M (x) = 0 $ and hence
$ \rho^\chi_* (\xi) \alpha_x = 0 $ from (\ref{Cartan}),
which implies that 
$ \rho^\chi(g) \alpha_x = \alpha_x $ for $ g \in G_x $
if $ G_x $ is connected.
Thus we conclude that $ \alpha_x \in ({\cal H}^\chi)^{G_x} $.

\subsection{Associated vector bundles}
So far we have discussed
equivariant functions and forms on the stratified bundle.
We now wish to define 
vector bundles associated with the stratified bundle,
like 
vector bundles associated with principal fiber bundles.

Let us define an equivalence relation $ \sim $ 
in $ M \times {\cal H}^\chi $ by the relation
$ ( x, v ) \sim  ( gx, \rho^\chi(g)v ) $ with $ g \in G $.
Let $ [x,v] $ denote the equivalence class 
with a representative $ ( x, v ) $.
Then the vector bundle associated 
with the stratified bundle $ ( M, G, Q, \pi, {\cal T}(M) ) $
by a representation $ ( {\cal H}^\chi, \rho^\chi ) $ 
is defined to be the quotient space 
\begin{equation}
	M \times_{\rho^\chi} {\cal H}^\chi
	:=
	\left(
	\, \coprod_{x \in M} 
	( \{ x \} \times ({\cal H}^\chi)^{G_x} ) 
	\right) / \sim.
	\label{associated vector bundle}
\end{equation}
The projection map
$ M \times {\cal H}^\chi \to M $ naturally induces
a projection map 
$ \pi_{\rho^\chi} : M \times_{\rho^\chi} {\cal H}^\chi \to Q $; 
$ [ x,v ] \mapsto \pi(x) $.
{}Further, each point $ x \in M $ defines an isomorphism
$ \tilde{x}: ({\cal H}^\chi)^{G_x} \to 
\pi_{\rho^\chi}^{-1} ( \pi (x) ) 
= {\cal O}_x \times_{\rho^\chi} {\cal H}^\chi $
by $ v \mapsto \tilde{x} (v) := [x,v] $. 
Note that, for each stratum $Q_{\tau}$, 
$ \pi_{\rho^\chi}^{-1} ( Q_\tau ) $ is
a vector bundle over $ Q_\tau $
with fiber $ ({\cal H}^\chi)^{G_\tau} $, so that 
one has $ M \times_{\rho^\chi} {\cal H}^\chi 
= \coprod_\tau \pi_{\rho^\chi}^{-1} ( Q_\tau ) $. 
In this sense, we may call 
$ M \times_{\rho^\chi} {\cal H}^\chi $ 
a {\it stratified vector bundle}. 
However, we will refer to it as the {\it associated vector bundle} 
for simplicity. 
Moreover, each fiber $ \pi_{\rho^\chi}^{-1} (q) $ at $ q \in Q $
inherits an inner product from $ {\cal H}^{\chi} $; 
for $ \eta, \eta' \in \pi_{\rho^\chi}^{-1}(q) $, 
the inner product $ \bra \eta, \eta' \ket $ is defined to be
\begin{equation}
	\bra \eta, \eta' \ket
	:=
	\bra \tilde{x}^{-1} (\eta), \tilde{x}^{-1} (\eta') \ket,
	\label{fiberwise inner product}
\end{equation}
where the RHS is the inner product defined on $ {\cal H}^\chi $.
It is easy to verify that the RHS is independent of the choice of 
$ x \in \pi^{-1}(q) $.

A {\it section} of the associated vector bundle 
is a map
$ \sigma : Q \to M \times_{\rho^\chi} {\cal H}^\chi $ satisfying
$ \pi_{\rho^\chi} \circ \sigma = \mbox{id}_Q $.
An equivariant function $ \psi $ defines 
a section $ \psi^\flat $ of the associated vector bundle through 
$ \psi^\flat( \pi(x) ) = [ x, \psi (x) ] 
= ( \tilde{x} \circ \psi ) (x) $.
The $ \psi^\flat(q) $ is well-defined.
In fact,
$ \psi^\flat( \pi(gx) ) 
= [ gx, \psi(gx) ] 
= [ gx, \rho^\chi(g) \psi(x) ]
= [ x,  \psi(x) ] = \psi^\flat( \pi(x) ) $ for $ g \in G $. 
Conversely, a section $ \sigma $ defines an equivariant 
function $ \sigma^\# $
through 
$ \sigma^\# (x) = ( \tilde{x}^{-1} \circ \sigma \circ \pi ) (x) $.
Thus we can identify sections of the associated vector bundle 
with equivariant functions.
When the corresponding equivariant function $ \sigma^\# $ 
is differentiable
as an $ {\cal H}^\chi $-valued function on $ M $,
the section $ \sigma $ is called differentiable,
while $ Q $ is not a manifold in general.
The set of all of the differentiable sections of
$ M \times_{\rho^\chi} {\cal H}^\chi $ is denoted by
$ {\mit\Gamma} ( M \times_{\rho^\chi} {\cal H}^\chi ) $.
Thus the space of sections
$ {\mit\Gamma} ( M \times_{\rho^\chi} {\cal H}^\chi ) $
is in one-to-one correspondence with
the space of equivariant functions
$ C^\infty (M; {\cal H}^\chi)^G $.

To associate equivariant horizontal $k$-forms with certain sections,
we should set up another vector bundle.
The action of $ g \in G $ on the space of one-forms, 
$ (g^{-1})^* : T^*_x M \to T^*_{gx} M $,
induces the action on the space of horizontal $ k $-forms,
$ (g^{-1})^* : \wedge^k H^*_x \to \wedge^k H^*_{gx} $.
This action is extended to that on
the space of $ {\cal H}^{\chi} $-valued horizontal $k$-forms  
$ \wedge^k H^* \otimes {\cal H}^\chi $
by the linear map generated by
\begin{equation}
\phi \otimes v 
\in \wedge^k H^*_x \otimes {\cal H}^\chi 
\mapsto
\widetilde{\rho^\chi}(g) (\phi \otimes v)
:= (g^{-1})^* \phi \otimes \rho^\chi(g) v 
\in \wedge^k H^*_{gx} \otimes {\cal H}^\chi. 
\label{ex-action}
\end{equation}
We then take a subspace of invariant vectors
\begin{equation}
	( \wedge^k H^*_x \otimes {\cal H}^\chi )^{G_x}
	:=
	\{
	\zeta \in \wedge^k H^*_x \otimes {\cal H}^\chi \, | \,
	\widetilde{\rho^\chi}(g) \zeta = \zeta, \,
	\forall g \in G_x
	\}
	\label{invariant subspace}
\end{equation}
and define an associated vector bundle, 
like (\ref{associated vector bundle}), through
\begin{equation}
	M \times_{\widetilde{\rho^\chi}} 
	( \wedge^k H^* \otimes {\cal H}^\chi )
	:=
	\left(
	\, \coprod_{x \in M}
	(
	\{ x \} \times
	( \wedge^k H^*_x \otimes {\cal H}^\chi )^{G_x}
	\right) / \sim,
	\label{complicated bundle}
\end{equation}
where the equivalent relation $ \sim $ in
$ M \times ( \wedge^k H^* \otimes {\cal H}^\chi ) $
is defined as $ (x, \zeta) \sim (gx, \widetilde{\rho^\chi}(g)\zeta) $.
A projection map
$ \pi_{\widetilde{\rho^\chi}} : 
M \times_{\widetilde{\rho^\chi}} ( \wedge^k H^* \otimes {\cal H}^\chi )
\to Q $ is defined naturally as $ [x,\zeta] \mapsto \pi(x) $.
The space of smooth sections 
$ {\mit\Gamma} 
( M \times_{\widetilde{\rho^\chi}} 
( \wedge^k H^* \otimes {\cal H}^\chi ) ) $
is also in one-to-one correspondence with
the space of equivariant horizontal forms
$ {\mit\Omega}_H^k (M; {\cal H}^\chi)^G $.

\subsection{Covariant derivative}
Covariant derivatives of $k$-forms on the stratified bundle 
can be defined like those on the principal fiber bundle.
{\it Covariant derivation} is a linear map
$ D : {\mit\Omega}_H^k     ( M; {\cal H}^\chi )^G 
 \to  {\mit\Omega}_H^{k+1} ( M; {\cal H}^\chi )^G $ defined through
\begin{equation}
	D \alpha ( X_1, \cdots, X_{k+1} )
	=
	d \alpha ( X_1^H, \cdots, X_{k+1}^H ),
	\qquad
	X_i \in T_x M, \: i=1, \cdots, k+1,
	\label{covariant derivative}
\end{equation}
for $ \alpha \in {\mit\Omega}_H^k ( M; {\cal H}^\chi )^G $,
where $ X_i^H $ is the horizontal component of $ X_i = X_i^V + X_i^H $.

The covariant derivative $ D \alpha $ can be expressed
by using the connection form $ \omega $.
To any $ X^V \in V_x $, there corresponds an element $ \xi \in \g $
such that $ X^V = \xi_M (x) $ uniquely modulo $ \g_x $.
By the definition of the connection form $ \omega $,
we then have $ \xi \equiv \omega( X^V ) \equiv \omega(X) $
$ (\mbox{mod} \, \g_x) $.
With the help of (\ref{Cartan}), one has
$ i( X^V )   d \alpha 
= i( \xi_M ) d \alpha
= \rho^\chi_*( \xi ) \alpha
= \rho^\chi_*( \omega(X) ) \alpha $.
The last equality holds well in spite of 
the ambiguity in the value of $ \omega(X) $
since $ \rho^\chi_*( \zeta ) \alpha_x = 0 $ for $ \zeta \in \g_x $.
Putting together these equations results in
\begin{eqnarray}
	&   &
	D \alpha ( X_1, \cdots, X_{k+1} )
	\nonumber
	\\
	& = &
	d \alpha 
	( X_1 - X_1^V, \cdots, X_{k+1} - X_{k+1}^V )
	\nonumber
	\\
	& = &
	d \alpha ( X_1, \cdots, X_{k+1} )
	+
	\sum_{i=1}^{k+1}
	(-1)^i
	\rho^\chi_* ( \omega(X_i) )
	\alpha 
	( X_1, \cdots, \widehat{X_i}, \cdots, X_{k+1} ),
\end{eqnarray}
where $ \widehat{X_i} $ means the removing of $ X_i $.
Thus our result is expressed as
\begin{equation}
	D \alpha 
	= d \alpha - \rho^\chi_* ( \omega ) \wedge \alpha.
	\label{D}
\end{equation}
We should note that smoothness of the covariant derivative
is ensured only within each stratum $ M_\tau $.

\section{Reduction of the Laplacian}
\subsection{Reduced Laplacians}
We have characterized the Hilbert space of the reduced quantum 
system and studied the geometric structure of the manifold 
which is underlying the Hilbert space.
Now we turn our attention to the Hamiltonian 
acting on the Hilbert space $ L_2(M) $. 
As was anticipated, the Hamiltonian $ H $ we treat 
takes the form 
$ H = \dfrac{1}{2} \Delta_M + V $
with the Laplacian $ \Delta_M $ and the potential $ V $. 

{}First we give a precise definition of the Laplacian.
We have been working with the $ G $-manifold $ M $
equipped with the $ G $-invariant measure $ \mu_M $.
We assume that $ M $ has no boundary.
In what follows, we make another assumption on $ M $;
$ M $ is assumed to be oriented
and equipped with a Riemannian metric $ g_M $
which is invariant under the action of $ G $.
The metric $ g_M $ induces a volume form $ v_ M $,
which is also invariant under the action of $ G $.
We assume also that
$ \mu_M $ is the measure associated with the volume form $ v_M $.
Let $ C^\infty_c (M) $ be
the set of all the $ C^\infty $ functions on $ M $
with compact support.
Then the Laplacian 
$ \Delta_M : C^\infty_c (M) \to C^\infty_c (M) $
is defined through
\begin{equation}
	\int_M || df (x) ||^2_{g_M} \, v_M 
	=
	\int_M f(x) ( \Delta_M f )(x) \, v_M,
	\label{Laplacian}
\end{equation}
where $ || df (x) ||^2_{g_M} $ denotes 
the norm of $ T^*_x M $ induced by the metric $ g_M $.
We note that the domain of $ \Delta_M $ is extended in $ L_2(M) $
to make $ \Delta_M $ a self-adjoint operator.
Of course, in order that this be the case, $ M $ has to be 
assumed to be complete. 
Since both the metric $ g_M $ 
and the volume $ v_M $ are $ G $-invariant,
the Laplacian $ \Delta_M $ is also $ G $-invariant;
namely, $ \Delta_M $ commutes with $ U(g) $ for any $ g \in G $.

Next we turn our attention to the potential energy $ V(x) $.
It is a smooth function $ V : M \to \R $
acting on $ f \in L_2(M) $ by
multiplication as $ ( Vf ) (x) := V(x) f(x) $.
We assume that $ V $ is also $ G $-invariant;
$ V(gx) = V(x)$ for any $ g \in G $ and $ x \in M $.
Thus the action of $ V $ also commutes with $ U(g) $.

Since each term of the Hamiltonian 
$ H = \dfrac{1}{2} \Delta_M + V $
commutes with $ U(g) $ for any $ g \in G $, 
we can apply the decomposition (\ref{decomposition}) 
and the commutativity (\ref{commutativity})
to both $ \Delta_M $ and $ V $ separately.
Hence, $ \Delta_M $ and $ V $ act as
$ ( \mbox{id}_{{\cal H}^{\chi}} \otimes \Delta_M ) $ and 
$ ( \mbox{id}_{{\cal H}^{\chi}} \otimes V ) $ on the reduced 
Hilbert space
$ ( {\cal H}^\chi \otimes L_2(M) )^G \cong L_2(M;{\cal H}^\chi)^G $. 
The Laplacian $ (\mbox{id}_{{\cal H}^{\chi}} \otimes \Delta_M) $ 
with the domain restricted to
$ ( {\cal H}^\chi \otimes L_2(M) )^G \cong L_2(M;{\cal H}^\chi)^G $
is denoted by $ \Delta^\chi $,
and is called a {\it reduced Laplacian}.
The reduced Laplacian $ \Delta^\chi $ is also a self-adjoint operator.
Then for an equivariant function
$ \psi \in C^\infty_c (M; {\cal H}^\chi)^G $,
the defining equation of the Laplacian $ \Delta^\chi $
takes the form
\begin{equation}
	\int_M
	|| d \psi(x) ||^2_{g_M} \, v_M
	=
	\int_M
	\bra \psi(x), (\Delta^\chi \psi)(x) \ket \, v_M,
	\label{Laplacian on equivariant function}
\end{equation}
where in the LHS
$ || d \psi(x) ||^2_{g_M} $ denotes
the norm of $ {\cal H}^\chi \otimes T^*_x M $
induced from the metric $ g_M $.

\subsection{Rotational and vibrational energy operators}
To make further study of $ \Delta^\chi $,
we make intensive use of
the vertical-horizontal decomposition $ T_x M = V_x \oplus H_x $
introduced in Sec. III. 
We have not chosen a specific connection yet.
Now we fix the connection by demanding the orthogonality
$ V_x \perp H_x $ with respect to the metric $ g_M(x) $.
Then, the $ G $-invariance of the metric ensures 
that $ g_* H_x = H_{gx} $,
and hence a unique connection is determined.
Since the set of maximum points $ M_\mu $ is 
an open dense subset of $ M $ as noticed previously 
and since the connection is smooth when restricted to $ M_\mu $,
the Laplacian on $ M_\mu $ will be smoothly decomposed 
into two, vertical and horizontal components,
by the use of the connection.

According to the orthogonal decomposition 
$ T^*_x M = H_x^* \oplus V_x^* $,
the integrand of the LHS of 
(\ref{Laplacian on equivariant function}) is also written as
\begin{equation}
	|| d \psi(x) ||^2_{g_M}
	= 
	|| (d \psi(x))_H ||^2_{g_M} +
	|| (d \psi(x))_V ||^2_{g_M}.
	\label{Laplacian decomposed}
\end{equation}
By the definition of 
the covariant derivative (\ref{covariant derivative})
and the expression (\ref{D}) in terms of the connection form,
the above equation is put in the form 
\begin{eqnarray}
	|| d \psi(x) ||^2_{g_M}
	& = &
	|| ( d - \rho^\chi_* (\omega_x) ) \psi(x) ||^2_{g_M} +
	||       \rho^\chi_* (\omega_x)   \psi(x) ||^2_{g_M}  
	\nonumber \\
	& = &
	|| D \psi(x) ||^2_{g_M} +
	|| \rho^\chi_* (\omega_x) \psi(x) ||^2_{g_M}.
	\label{Laplacian decomposed 2}
\end{eqnarray}
The second term of the RHS of (\ref{Laplacian decomposed 2}) 
proves to be expressed as
\begin{eqnarray}
	|| \rho^\chi_* (\omega_x) \psi(x) ||^2_{g_M}
	& = &
	g_M^{-1}
	\bra
		\rho^\chi_*(\omega_x) \psi(x),
		\rho^\chi_*(\omega_x) \psi(x)
	\ket
	\nonumber \\
	& = &
	- g_M^{-1}
	\bra
		\psi(x),
		\rho^\chi_*(\omega_x) \otimes
		\rho^\chi_*(\omega_x) \psi(x)
	\ket
	\nonumber \\
	& = &
	\bra
		\psi(x),
		{\mit\Lambda}^\chi_x \, \psi(x)
	\ket,
	\label{vertical Laplacian}
\end{eqnarray}
where we have to make remarks on the notations used; 
the $ g_M^{-1}\bra\;,\,\ket $ denotes the inner product 
on $ T^*M\otimes {\cal H}^{\chi} $, 
the product $ \rho^\chi_*(\omega_x) \otimes \rho^\chi_*(\omega_x) $
is to be understood as a tensor product in 
$ T^*M \otimes T^*M \otimes \mbox{End}{\cal H}^{\chi} $ and 
the $ {\mit\Lambda}^{\chi}_x $ is defined by 
\begin{equation}
	{\mit\Lambda}^\chi_x := 
	- g_M^{-1} (x)\circ
          ( \rho^\chi_* (\omega_x )\otimes \rho^\chi_* (\omega_x ))
	 \in \mbox{End} \,  {\cal H}^\chi 
	\label{Lambda}
\end{equation}
with $ g_M^{-1}(x) $ taken as the inner product on 
$ T^*M \otimes T^*M $. 
We notice also that at the second line of 
(\ref{vertical Laplacian}),  
we have used the fact that 
$   \bra \rho^\chi_*(\xi) v, v' \ket 
= - \bra v, \rho^\chi_*(\xi) v' \ket $
for any $ \xi \in \g $ and for any $ v, v' \in {\cal H}^\chi $.
The equivariance of the connection form  (\ref{property2}) 
and the invariance of the metric are put together to imply that
\begin{equation}
 {\mit\Lambda}^\chi_{gx} = 
\rho^\chi(g) \, {\mit\Lambda}^\chi_x \, \rho^\chi(g^{-1}) .
\end{equation}
Then we observe that the operator
$ {\mit\Lambda}^\chi $ acting on
$ L_2 (M; {\cal H}^\chi) $ through 
$ ({\mit\Lambda}^\chi \, \psi) (x) 
:= {\mit\Lambda}^\chi_x \, \psi (x) $
leaves $ L_2 (M; {\cal H}^\chi)^{G} $ invariant, so that 
one obtains 
$ \mit\Lambda_x^{\chi} \in \mbox{End}({\cal H}^{\chi})^{G_x} $. 
We can put the $ {\mit\Lambda}^{\chi} $ in another form. 
Since the inner product $ g_M(x) : T_xM \otimes T_xM \to \R $ 
gives rise to an isomorphism $ \widehat{g}_M(x) : T_xM \to T^*_x M $,
its inverse
$ \widehat{g}_M^{-1}(x) : T_x^* M \to T_xM $ induces 
an inner product on the cotangent space 
$ g_M^{-1}(x) : T_x^* M \otimes T_x^* M \to \R $ 
in the dual manner. 
Hence $ g^{-1}_M $ is viewed as a symmetric tensor field
$ g_M^{-1} : M \to TM \otimes TM $. 
In terms of $ g^{-1}_M $ along with 
the connection form $ \omega_x : T_x M \to \g / \g_x $ 
and the representation of Lie algebra
$ \rho^\chi_* : \g \to \mbox{End} \, {\cal H}^\chi $,
the tensor field $ \mit\Lambda^{\chi} : 
M \to \mbox{End} {\cal H}^{\chi} $ takes the form 
\begin{equation}
	{\mit\Lambda}^\chi := 
	- ( \rho^\chi_* \otimes \rho^\chi_* ) \circ 
	  ( \omega \otimes \omega ) \circ g_M^{-1}.
	\label{Lambda tensor}
\end{equation}
If $ \rho^\chi $ is not a trivial representation
and if $ G $ acts on $ M $ nontrivially,
then $ {\mit\Lambda}^\chi $ is a positive definite operator.
We call the $ {\mit\Lambda}^\chi $ 
the {\it rotational energy operator},
the reason for which comes from molecular mechanics 
with $ G = SO(3) $. 
In fact, when applied to molecular mechanics, the quantity (4.5) 
is interpreted as the rotational energy density. 

We proceed to study the first term of the RHS of 
(\ref{Laplacian decomposed 2}). 
To this end, we have to make a review of 
{\it Hodge's star operator} 
$ *_M : {\mit\Omega}^k     (M) 
\to     {\mit\Omega}^{m-k} (M) $, 
which is defined through
\begin{equation}
	\alpha \wedge *_M \beta
	=
	\bra \alpha, \beta \ket_{g_M} \, v_M,
	\qquad
	\alpha, \beta \in {\mit\Omega}^k (M),
	\label{star}
\end{equation}
where $ \bra \alpha, \beta \ket $ in the RHS is
the inner product on $ \wedge^k T^*_x M $ 
defined by the metric $ g_M $.
Let $ \{ e_1, \cdots, e_m \} $ be a local orthonormal frame field
on $ M $ with respect to the metric $ g_M $.
Then the star operator $ *_M $ is explicitly given by
\begin{equation}
	*_M \alpha 
	=
	\frac{1}{k!}
	\sum_{j_1, \cdots, j_k = 1}^m
	( i(e_{j_1}) \cdots i(e_{j_k}) \alpha )
	\,
	( i(e_{j_1}) \cdots i(e_{j_k}) v_M ).
	\label{explicit star}
\end{equation}
It is easily verified that
$ *_M *_M \alpha = (-1)^{ k(m-k) } \alpha $.
The defining equation (\ref{Laplacian}) of the Laplacian $ \Delta_M $
is then rewritten as
\begin{eqnarray}
	\int_M || df (x) ||^2_{g_M} \, v_M 
	& = &
	\int_M df \wedge *_M df
	\nonumber \\
	& = &
	\int_M d ( f *_M df ) 
	- \int_M f \, d *_M df
	\nonumber \\
	& = &
	- \int_M f ( *_M^{-1} d *_M df ) \, v_M,
	\label{Laplacian 2}
\end{eqnarray}
where we have used Stokes' theorem 
to eliminate the first term on the second line.
Then the Laplacian takes the form 
\begin{equation}
	\Delta_M f 
	= - *_M^{-1} d *_M d f
	= - *_M      d *_M d f.
	\label{Laplacian 3}
\end{equation}
In terms of local coordinates $ (x^1, \cdots, x^m) $ of $ M $,
the metric and the volume form are expressed as 
$ g_M = \sum_{i,j} g_{Mij} (x) \, dx^i \otimes dx^j $ and
$ v_M 
= v_M (x) \, dx^1 \wedge \cdots \wedge dx^m 
= \sqrt{ \det g_{Mij}(x) } \, dx^1 \wedge \cdots \wedge dx^m $,
respectively.
The Laplacian $ \Delta_M $ is then expressed as
\begin{equation}
	\Delta_M f
	=
	- \frac{1}{v_M (x)} 
	\sum_{i,j=1}^m
	\frac{\partial}{\partial x^i}\Bigl(
	v_M (x) (g_M^{-1})^{ij} (x)
	\frac{\partial f}{\partial x^j}\Bigr),
	\label{Laplacian coordinate}
\end{equation}
as is well-known.

We examine (\ref{Laplacian 3}) more closely,
using the horizontal-vertical decomposition 
$ T_x M  = H_x \oplus V_x $.
The measure $ \mu_M $ of $ M $
projects to a measure $ \mu_Q $ of $ Q $ through 
$ \pi : M \to Q $;
\begin{equation}
	\int_Q \varphi(q) d \mu_Q (q)
	:=
	\int_M ( \varphi \circ \pi )(x) d \mu_M (x),
	\qquad
	\varphi \in C^0(Q).
	\label{muQ}
\end{equation}
In what follows, we restrict ourselves to 
the maximum stratum $ M_\mu $,
which is an open and dense subset of $ M $. 
Let $ v_Q $ be a volume form on $ Q_\mu $ associated with
the measure $ \mu_Q $.
We can define a Riemannian metric $ g_Q $ on $ Q_\mu $
through $ \pi^* g_Q = g_M|H $,
where $ g_M|H $ denote the restriction of
the metric $ g_M : T_x M \times T_x M \to \R $
to the horizontal subspace; $ g_M|H : H_x \times H_x \to \R $.
Note that the definition of $ g_Q $ is independent of the choice 
of $ x \in \pi^{-1}(\pi(x))$, because of the $G$-invariance of 
$ g_M $. 
The map $ \pi : (M_\mu, g_M) \to (Q_\mu, g_Q) $ then becomes
a Riemannian submersion. 
It is to be noted that 
the volume form $ v_Q $ does not coincides with 
the volume form induced from the metric $ g_Q $.
We denote the set of all the horizontal and the vertical 
$ k $-forms on $ M_\mu $ by 
\begin{eqnarray}
	{\mit\Omega}^k_H (M_\mu) 
	& := &
	\{ 
		\alpha \in {\mit\Omega}^k (M_\mu) \, | \,
		i(X) \alpha = 0, \, 
		\forall X \in V_x, \, 
		\forall x \in M_\mu 
	\},
	\label{H-form}
	\\
	{\mit\Omega}^k_V (M_\mu) 
	& := &
	\{ 
		\alpha \in {\mit\Omega}^k (M_\mu) \, | \,
		i(X) \alpha = 0, \, 
		\forall X \in H_x, \, 
		\forall x \in M_\mu 
	\},
	\label{V-form}
\end{eqnarray}
respectively. 
Note that we have already put 
$ \dim M = m $, $ \dim Q_\mu = n = m-p $.
We define 
the horizontal and the vertical volume forms, 
$ v_H \in {\mit\Omega}^n_H (M_\mu) $ 
and 
$ v_V \in {\mit\Omega}^p_V (M_\mu) $, 
through 
$ v_H = \pi^* v_Q $ and $ v_M = v_H \wedge v_V $, respectively. 
It appears that the forms $ v_H $ and $ v_V $ are uniquely 
determined and $ G $-invariant. 

Moreover, it can be shown that both $ v_H $ and $ v_V $ 
are closed forms. 
It is easy to verify that $ v_H $ is closed; 
$ d v_H 
= d ( \pi^* v_Q )
= \pi^* ( d v_Q )
= 0 $,
since $ v_Q $ is a top form of $ Q_\mu $. 
To prove that $ d v_V = 0 $, 
we use a local trivialization over an open set $ W \subset Q_\mu $;
$ \pi^{-1} (W) \subset M_\mu $ is identified with $ W \times F $,
where $ F := G/G_\mu $ is the maximum orbit.
The trivialization induces a surjective map
$ \pi_F : \pi^{-1}(W) \to F $ which is $ G $-equivariant,
that is, 
$ \pi_F(gx) = g \pi_F(x) $ 
for each $ g \in G $ and $ x \in \pi^{-1}(W) $.
A restriction of the map $ \pi_F $ to each fiber
gives a diffeomorphism
$ \pi^{-1}(q) \cong F $ for each $ q \in W $.
Let $ v_F $ be a $ G $-invariant volume form on $ F $
which is normalized as $ \int_F v_F = 1 $.
Then $ v_F $ is uniquely determined.
Now the definition (\ref{muQ})
of the measure $ \mu_Q $ is put in the form 
\begin{eqnarray}
	\int_{W} \varphi(q) \, v_Q 
	& = &
	\int_{\pi^{-1}(W)} (\varphi \circ \pi)(x) \, v_M
	= 
	\int_{\pi^{-1}(W)} (\varphi \circ \pi)(x) \, v_H \wedge v_V
	\nonumber \\
	& = &
	\int_{W} \varphi(q) \, v_Q
	\left( \int_{\pi^{-1}(q)} v_V \right).
	\label{calculation of v_V}
\end{eqnarray}
Since $ \varphi $ is arbitrary, this equation implies that 
the volume form $ v_V $ restricted to each fiber $ \pi^{-1}(q) $
is also normalized as $ \int_{\pi^{-1}(q)} v_V = 1 $ 
for each $ q \in Q_\mu $, so that
$ v_V $ is a $ G $-invariant normalized volume form on each fiber
$ \pi^{-1} (q) $. 
It then follows that $ v_V = \pi_F^* v_F $.
As a consequence, one has
$ d v_V = \pi_F^* ( d v_F ) = 0 $, 
since $ v_F $ is a top form of $ F $.

Using $ v_H $ and $ v_V $,
we define 
the horizontal and the vertical star operators 
$ *_H : {\mit\Omega}^k_H (M_\mu) \to {\mit\Omega}^{n-k}_H (M_\mu) $
and
$ *_V : {\mit\Omega}^k_V (M_\mu) \to {\mit\Omega}^{p-k}_V (M_\mu) $
through
\begin{eqnarray}
	*_M \alpha & = & ( *_H \alpha ) \wedge v_V,
	\qquad
	\alpha \in {\mit\Omega}^k_H (M_\mu) \, (k \le n),
	\label{H star}
	\\
	{}*_M \beta & = & (-1)^n \, v_H \wedge *_V \beta,
	\qquad
	\beta \in {\mit\Omega}^k_V (M_\mu) \, (k \le p),
	\label{V star}
\end{eqnarray}
respectively. 
According to the decomposition $ T_x^* M = H_x^* \oplus V_x^* $, 
we break up $ df $ into $ df = (df)_H + (df)_V $. 
Then $ *_M df $ is accordingly expressed as 
\begin{equation}
	*_M df
	=
	*_H (df)_H \wedge v_V + (-1)^n v_H \wedge *_V (df)_V.
\end{equation}
Since $ dv_H = 0 $ and $ dv_V = 0 $, we obtain 
\begin{equation}
	d *_M df
	=
	( d *_H (df)_H ) \wedge v_V 
	+ v_H \wedge ( d *_V (df)_V ).
\end{equation}
Thus Eq.(\ref{Laplacian 3}) is expressed as
\begin{equation}
	- \Delta_M f 
	=
	*_M^{-1} d *_M df
	=
	*_H^{-1} ( d *_H (df)_H )_H +
	*_V^{-1} ( d *_V (df)_V )_V,
	\label{Laplacian 4}
\end{equation}
which means that the Laplacian $ \Delta_M $ is 
decomposed into horizontal and vertical components. 

The above argument can be extended to 
$ {\cal H}^\chi \otimes {\mit\Omega}^k(M_{\mu})
\cong {\mit\Omega}^k(M_{\mu}; {\cal H}^\chi) $ and to
$ {\mit\Omega}^k_H (M_\mu; {\cal H}^\chi)^G $ straightforwardly; 
the star operators $ *_M $ and $ *_H $ are extended to 
be applicable to $ {\cal H}^{\chi} $-valued forms on $ M_{\mu} $ 
and to $ {\cal H}^{\chi} $-valued horizontal forms on $ M_{\mu} $, 
respectively. 
Hence, for an equivariant function 
$ \psi \in C^{\infty}_c(M_{\mu};{\cal H}^{\chi})^G $, 
Eq.(\ref{Laplacian 4}) gives rise to 
\begin{equation}
	- \Delta^{\chi} \psi 
	=
	*_M^{-1} d *_M d\psi
	=
	*_H^{-1} ( d *_H (d\psi)_H )_H +
	*_V^{-1} ( d *_V (d\psi)_V )_V.
	\label{Laplacian 5}
\end{equation}

{}For $ \psi \in C^\infty_c (M_{\mu}; {\cal H}^\chi)^G $,
we have $ (d \psi)_H = D \psi $ by the definition of 
the covariant derivation (\ref{covariant derivative}).  
In view of the first term of the RHS of (\ref{Laplacian 5}), 
we are led to the definition of the adjoint operator 
$ D^\dagger : {\mit\Omega}_H^{k+1} (M_\mu; {\cal H}^\chi)^G $
$ \to $     $ {\mit\Omega}_H^k     (M_\mu; {\cal H}^\chi)^G $ 
of $ D $; 
\begin{equation}
	D^\dagger := - *_H^{-1} D *_H = - (-1)^{k(n-k)} *_H D *_H.
	\label{adjoint D}
\end{equation}
By using 
(\ref{Laplacian on equivariant function}),
(\ref{Laplacian decomposed 2}),
(\ref{vertical Laplacian}),
(\ref{Laplacian 4}), and
(\ref{adjoint D}), 
we accomplish the decomposition of the Laplacian $ \Delta^\chi $
into the horizontal and the vertical components;
\begin{equation}
	\Delta^\chi \psi
	=
	D^\dagger D \psi + {\mit\Lambda}^\chi \psi,
	\qquad
	\psi \in C^\infty_c (M_{\mu}; {\cal H}^\chi)^G.
	\label{Laplacian chi}
\end{equation}
We call the $ D^\dagger D $ 
the {\it vibrational energy operator}
for the reason that the integral (4.2) is interpreted as the 
vibrational energy 
when our general formalism is applied to molecular mechanics.

As was noted above,
the smoothness of the horizontal-vertical decomposition is ensured
only in the open dense subset $ M_\mu \subset M $,
so that the RHS of (\ref{Laplacian chi}) makes sense only 
in $ M_\mu $.
However, $ \Delta^\chi $ is actually 
a self-adjoint operator on $ L_2(M;{\cal H}^{\chi}) $ by definition, 
so that the LHS of (4.2) holds for $ \psi $ defined throughout $ M $. 
Hence we may expect that some boundary condition arises on 
$ \partial M_\mu = M_{\mbox{\scriptsize sing}} $ 
to make $ D^\dagger D + {\mit\Lambda}^{\chi} $
into a self-adjoint operator on $ L_2(M;{\cal H}^{\chi}) $. 
Since the requirement imposed on $ \psi $ is
that $ \psi $ is to be equivariant, we obtain the boundary 
condition on $ \partial M_{\mu} $, 
\begin{equation}
	\rho^\chi (g) \psi (x) = \psi (x),
	\qquad
	g \in G_x, \; x \in \partial M_{\mu}
	\label{boundary condition1}
\end{equation}
or, in the form of derivative, 
\begin{equation}
	\rho^\chi_* (\xi) \psi (x) = 0,
	\qquad
	\xi \in \g_x, \; x \in \partial M_{\mu}.
	\label{boundary condition2}
\end{equation}
In case of $ G_x\neq \{e\} $ for the maximum orbit type $ \mu $, 
the equivariant functions are, of course, subject to the 
condition $ \rho^\chi (g) \psi (x) = \psi (x) $ for $ g \in G_x,\;
x \in M_{\mu} $. 
The condition (\ref{boundary condition1}) 
or (\ref{boundary condition2}) 
states that at singular points the equivariant functions 
are subject to a stronger condition since $ \mbox{dim} G_x $ rises 
up at singular points. 

\subsection{Angular momentum and inertia tensor}
Now we wish to introduce the angular momentum and 
the inertia tensor, which are closely related with 
the connection form and the rotational energy operator 
$ \mit\Lambda^{\chi} $. 

The {\it angular momentum} is
a map $ L : T^* M \to \g^* $; 
which is defined through
\begin{equation}
	\bra L_x(p), \xi \ket
	:=
	\bra p, \theta_x (\xi) \ket,
	\qquad
	p \in T^*_x M, \; \xi \in \g,
	\label{angular momentum map}
\end{equation}
where $ \bra \cdot, \cdot \ket $'s 
in the LHS and in the RHS denote the pairing between 
$ \g^* $ and $ \g $ and that between 
$ T^*_x M $ and  $ T_x M $, respectively, and 
$ \theta_x(\xi)(=\xi_M(x)) $ is the infinitesimal 
generator induced by $ \xi \in \g $. 
The angular momentum $ L : T^* M \to \g^* $ is 
a typical example of momentum maps due to 
Marsden and Souriau\cite{Souriau}.
By the use of the isomorphism 
$ \widehat{g}_M : TM \to T^* M $, 
the angular momentum can be rewritten as 
the map $ \widehat{L} := L \circ \widehat{g}_M : TM \to \g^* $;
$ v \mapsto \widehat{L}_x (v) $, 
which is expressed as 
\begin{equation}
	\bra \widehat{L}_x(v), \xi \ket 
	:=
	g_M ( v, \theta_x (\xi) ),
	\qquad
	v \in T_x M, \; \xi \in \g. 
	\label{dual angular momentum map}
\end{equation}
Namely, $ \widehat{L} $ is a $ \g^* $-valued one-form on $ M $, 
which we call the angular momentum form.

The {\it inertia tensor} is a tensor field 
$ I : M \to \g^* \otimes \g^* $; $ x \mapsto I_x $, 
which is defined through
\begin{equation}
	I_x ( \xi, \zeta ) 
	:= 
	g_M ( \theta_x(\xi), \theta_x(\zeta) ),
	\qquad
	\xi, \; \zeta \in \g.
	\label{inertia tensor}
\end{equation}
On account of $ \theta_{gx} (\Ad_g \xi) = g_* ( \theta_x (\xi) ) $ 
and of $ g^* \, g_M = g_M $ for any $ g \in G $,
the inertia tensor transforms according to
\begin{equation}
	I_{gx} ( \Ad_g \xi, \Ad_g \zeta ) 
	= 
	I_x  ( \xi, \zeta ).
\end{equation}
In other word, the map
$ I : M \to \g^* \otimes \g^* $
is equivariant; 
$ g^* I = ( \Ad_{g^{-1}}^* \otimes \Ad_{g^{-1}}^* ) I $.
{}For an arbitrary $ x \in M $ fixed, the quadratic form 
$ I_x : \g \otimes \g \to \R $
can be regarded as a map 
$ \widehat{I}_x : \g \to \g^* $,
which has 
$ \Kernel \, \widehat{I}_x = \g_x $ and 
$ \Image  \, \widehat{I}_x 
= \{ \phi \in \g^* \, | \, 
     \bra \phi, \xi \ket = 0, \, \forall \xi \in \g_x \}
\cong (\g/\g_x)^* $.
Then it can give rise to  
an isomorphism 
$ \widetilde{I}_x : \g/\g_x \isoarrow (\g/\g_x)^* $.
Hence, there exists the inverse 
$ (\widetilde{I}_x)^{-1} : (\g/\g_x)^* \isoarrow \g/\g_x $, 
which is identified with a quadratic form 
$ (\widetilde{I}_x)^{-1} : (\g/\g_x)^* \otimes (\g/\g_x)^* 
\to  \R $. 
The map $ \widehat{I}_x $ will be referred to as an inertia 
operator. 
The inertia operator is called the locked inertia tensor
by Simo et al.\cite{Simo}. 
The inertia operator was first introduced by Guichardet\cite{Guichardet}, 
and used in \cite{Tachibana-Iwai,Iwai3} to break up the total energy 
into the sum of rotational and vibrational energies.

{}From definition, the angular momentum, the inertia tensor, 
and the connection form turn out to be related by 
\begin{equation}
	\widehat{L} = \widehat{I} \circ \omega,
	\label{L=I omega}
\end{equation}
where each symbol is to be understood as follows: 
$ \widehat{L}_x : T_x M \to \g^* $,
$ \omega_x : T_x M \to \g/\g_x $, and
$ \widehat{I}_x : \g/\g_x \to \g^* $.
A proof of (\ref{L=I omega}) runs as follows:  
{}First, the identity
\begin{equation}
	\widehat{L} \circ \theta = \widehat{I},
	\label{L theta=I}
\end{equation}
can be proved by a straightforward calculation. In fact, 
for any $ \xi, \zeta \in \g $, we have
\begin{equation}
	\bra (\widehat{L} \circ \theta_x )(\xi), \zeta \ket
	=
	g_M ( \theta_x (\xi), \theta_x (\zeta) )
	=
	I_x ( \xi, \zeta )
	=
	\bra \widehat{I}_x (\xi), \zeta \ket.
\end{equation}
Next, from the identity (\ref{L theta=I}) it is deduced that
\begin{equation}
	\widehat{L} \circ \theta \circ \omega 
	= \widehat{I} \circ \omega.
	\label{L theta=I next}
\end{equation}
{}From the definition of the angular momentum form
(\ref{dual angular momentum map}), it can be shown that
$ \Kernel \, \widehat{L}_x = H_x $, so that Eq.(\ref{L=I omega}) 
holds on $ H_x $. 
Moreover, since 
$ \widetilde{\theta}_x \circ \omega_x : T_x M = V_x \oplus H_x \to V_x $
is a projection as was noted at (\ref{property3}), 
Eq.(\ref{L theta=I next}) shows that 
(\ref{L=I omega}) holds on $ V_x $. 
The proof is thus completed.
Since $ \mbox{Im}\widehat{L}_x \cong (\g/\g_x)^*$, 
we may rewrite (\ref{L=I omega}) as $\widehat{L} = 
\widetilde{I} \circ \omega$, so that the connection form 
is put in the form 
\begin{equation}
	\omega = \widetilde{I}^{-1} \circ \widehat{L}.
	\label{calculator of omega}
\end{equation}
This formula will be used to write out the connection 
form $ \omega $ in molecular mechanics. 

Owing to the definition of the inertia tensor, 
$ I_x = g_M \circ ( \theta_x \otimes \theta_x ) $, 
the metric $ g_M|V $ restricted to the vertical subspace 
takes the form 
\begin{equation}
	g_M|V
	= I \circ ( \omega \otimes \omega )
	= \bra \widehat{L}, \omega \ket,
	\label{rotational energy}
\end{equation}
where use has been made of $\widetilde{\theta}_x \circ \omega_x|V_x = 
\mbox{id}_{V_x}$, another expression of (\ref{property3}). 
Equation (\ref{rotational energy}) can be looked upon as describing 
the rotational energy in classical mechanics.
Since the reduced quadratic form 
$ \widetilde{I}_x 
= g_M \circ ( \widetilde{\theta}_x \otimes \widetilde{\theta}_x ) 
: \g/\g_x \otimes \g/\g_x \to \R $
is nondegenerate,
it has the inverse quadratic form 
$ (\widetilde{I}_x)^{-1} 
= g_M^{-1} \circ ( \omega_x^* \otimes \omega_x^* ):\;
(\g/\g_x)^* \otimes (\g/\g_x)^* \to \R $, 
which is expressed as a tensor field, 
$ x \mapsto (\widetilde{I}_x)^{-1} \in \g/\g_x \otimes \g/\g_x$, 
\begin{equation}
	( \widetilde{I}_x )^{-1}
	= ( \omega_x \otimes \omega_x ) \circ g_M^{-1}
	\label{inverse I}
\end{equation}
where 
$ g_M^{-1} $ is regarded as a symmetric tensor field
$ M \to TM \otimes TM $. 
{}From (\ref{Lambda tensor}) and (\ref{inverse I}), we obtain 
a formula to express $ {\mit\Lambda}^{\chi} $,  
in terms of the inertia tensor, as 
\begin{equation}
	{\mit\Lambda}^\chi = 
	- ( \rho^\chi_* \otimes \rho^\chi_* ) 
	\circ \widetilde{I}^{-1}.
	\label{Lambda and I}
\end{equation}

\subsection{Coordinate representation}
Now we wish to provide a coordinate representation of
the reduced Laplacian given in (\ref{Laplacian chi}).
Take local coordinates $ (q^1, q^2, \cdots, q^n ) $ 
on an open subset $ W $ of $ Q_\mu $.
Then the metric $ g_Q $ and the volume form $ v_Q $ 
take the form
$ g_Q = \sum_{i,j} g_{Qij}(q) \, dq^i \otimes dq^j $ and
$ v_Q = v_Q(q) \, dq^1 \wedge \cdots \wedge dq^n $, 
respectively.
{}For each $ x \in \pi^{-1}(W) \subset M_\mu $, 
we take a basis $ \{ \xi_1(x), \cdots, \xi_p(x) \} $
of $ \g/\g_x $, where we have used the same notation for 
elements of $ \g/\g_x $ as those for $ \g $ for simplicity. 
Then the components of the reduced inertia tensor 
$ \widetilde{I}_x $ are defined by
\begin{equation}
	( \widetilde{I}_x )_{\alpha \beta}
	:=
	\widetilde{I}_x ( \xi_\alpha(x), \xi_\beta(x) ),
	\qquad
	\alpha, \beta = 1, \cdots, p=\dim \g/\g_x,
\end{equation}
which gives a symmetric positive definite matrix 
of rank $ p $.
The components of its inverse are denoted by
$ ( \widetilde{I}_x^{-1} )^{\alpha \beta} $.
Let $ \sigma $ be a local section on $ W $
of the bundle $ \pi : M_\mu \to Q_\mu $,
that is, a differentiable map $ \sigma : W \to M_\mu $
such that $ \pi \circ \sigma = \mbox{id}_W $.
Then an equivariant function 
$ \psi \in C^\infty_c (M; {\cal H}^\chi)^G $
is pulled back to a function 
$ \sigma^* \psi : W \to {\cal H}^\chi $ 
of the coordinates $ (q^1, q^2, \cdots, q^n ) $.
Our aim is to obtain a coordinate expression of 
$  \sigma^* \Delta^\chi \psi $,
according to the decomposition
$ \Delta^\chi = D^\dagger D + {\mit\Lambda}^\chi $.
The covariant derivative (\ref{D}) with $ \alpha = \psi $
is expressed, in terms of $ (q^1, q^2, \cdots, q^n ) $, as
\begin{equation}
	( \sigma^* D \psi ) (q)
	=
	\sum_{i=1}^n
	\left( 
	\frac{\partial}{\partial q^i}
	- ( \rho^\chi_* \circ \omega \circ \sigma_* )
	\left( \frac{\partial}{\partial q^i} \right)
	\right)
	( \sigma^* \psi ) (q)
	d q^i.
	\label{pullbacked D}
\end{equation}
It then turns out from (\ref{adjoint D}), (\ref{Laplacian chi}), 
and (\ref{Lambda and I}) that 
\begin{eqnarray}
	( \sigma^* \Delta^\chi \psi ) (q)
	& = &
	- \frac{1}{v_Q(q)}
	\sum_{i,j=1}^n
	\left( 
	\frac{\partial}{\partial q^i}
	- ( \rho^\chi_* \circ \omega \circ \sigma_* )
	\left( \frac{\partial}{\partial q^i} \right)
	\right)
	\nonumber \\
	&&
	\qquad 
	v_Q(q)
	(g_Q^{-1})^{ij} (q)
	\left( 
	\frac{\partial}{\partial q^j}
	- ( \rho^\chi_* \circ \omega \circ \sigma_* )
	\left( \frac{\partial}{\partial q^j} \right)
	\right)
	( \sigma^* \psi ) (q)
	\nonumber \\
	& - &
	\! \! \!
	\sum_{\alpha, \beta=1}^p
	( \widetilde{I}^{-1}_{\sigma(q)} )^{\alpha \beta}
	( \rho^\chi_* \circ \xi_\alpha \circ \sigma)(q) 
	( \rho^\chi_* \circ \xi_\beta  \circ \sigma)(q) 
	( \sigma^* \psi ) (q).
	\label{calculated Laplacian}
\end{eqnarray}
It should be noted here that 
$ \omega(\sigma_*(\partial/\partial q^i)) $ 
and $ \rho^{\chi}_*(\omega(\sigma_*(\partial/\partial q^i))) $ 
denote a component of a ``gauge potential" and its representation as 
a matrix acting on ${\cal H}^{\chi} $, respectively. 
{}Further, $ \rho^{\chi}_*(\xi_{\alpha}(\sigma(q))) $ denote a 
matrix representation of the infinitesimal generator (or the 
angular momentum operator) induced by $ \xi_{\alpha}(x) \in \g $. 
This is one of our main results.

If there exists a global section $ \sigma : Q_\mu \to M_\mu $,
the fiber bundle $ \pi : M_\mu \to Q_\mu $ 
becomes a trivial bundle; $ M_\mu \cong Q_\mu \times (G/G_\mu) $,
and $ \sigma^* \psi $ becomes a smooth function 
over the entire domain $ Q_\mu $.
In this case, 
the procedure of reduction is nothing but separation of variables.
The reduction method is a generalization of
the method of separation of variables.

Equation  (\ref{calculated Laplacian}) makes sense
only in the maximum component, $ Q_{\mu} $, of the 
orbit space $ Q $. 
At a singular point $ q \in \partial Q_\mu $,
the rank of the inertia tensor $ I_{\sigma(q)} $ decreases abruptly.
As was noticed earlier, 
the equivariance condition provides the boundary condition 
imposed on $ \psi $ and hence on $ \sigma^* \psi $, 
which is put in the form 
\begin{equation}
	\rho^\chi_* ( \xi ) ( \sigma^* \psi ) (q) = 0,
	\qquad
	\xi \in \g_{\sigma(q)}.
	\label{boundary condition3}
\end{equation}
If $\g_{\sigma(q)}=\{0\} $ in $ M_{\mu} $, this imposes no 
condition on $ \sigma^*\psi $. 
At singular points $ q \in \partial Q_\mu $, 
the dimension of the isotropy algebra $ \g_{\sigma(q)} $ jumps up,
so that the value of $ \sigma^* \psi $
is more strongly restricted there.
The operator (\ref{calculated Laplacian}) should be accompanied by
the condition (\ref{boundary condition3}).
This is another one of our main results.

\subsection{Example: quantum mechanics on a plane}
Quantum mechanics in a two-dimensional Euclidean space
provides a simple but nontrivial example of 
the formulation constructed above.

{}First we take $ M = \R^2 $ and $ G = SO(2) $.
Then $ SO(2) $ acts on $ \R^2 $ in the usual manner. 
Let $ (r, \phi) $ be the polar coordinate of $ \R^2 $.
Then the orbit space becomes $ Q = \Rplus $, which is 
nothing but the radius coordinate $ r \ge 0 $. 
A point with $ r \ne 0 $ has a maximum orbit $ S^1 $.
The origin $ r = 0 $ has a singular orbit, that consists of
a single point $ \{ 0 \} $.
In this example we have
$ M_\mu = \R^2 - \{0\} $, $ \partial M_\mu = \{0\} $,
$ Q_\mu = \R_{>0} $, and $ \partial Q_\mu = \{0\} $.
Since $ \R^2 - \{0\} \cong \R_{>0} \times S^1 $, 
the fiber bundle $ \pi : M_\mu \to Q_\mu $ is trivial.
Then the reduction method 
becomes the method of separation of variables in this case.

The space $ \R^2 $ is equipped with the standard metric
$ g_M = dr^2 + r^2 d \phi^2 $
and the standard measure
$ d \mu_M = r dr d \phi $.
Then the projected measure of $ \Rplus $ is
$ d \mu_Q = 2 \pi r dr $.
At each maximum point $ x \in \R^2 $ with $ r \ne 0 $,
the vertical and horizontal subspaces are given by
\begin{equation}
	V_x = \R \frac{\partial}{\partial \phi},
	\qquad \quad
	H_x = \R \frac{\partial}{\partial r},
	\label{V and H of R^2}
\end{equation}
respectively.
At the singular point $ r = 0 $, one has 
$ V_0 = \{ 0 \} $ and $ H_0 \cong \R^2 $.
The metric on $ \R_{>0} $ is $ g_Q = dr^2 $, and thereby one obtains 
the Riemannian submersion $ \pi: M_\mu \to Q_\mu $ .
It should be noted that 
$ d \mu_Q = 2 \pi r dr $ does not coincide 
with the metric volume $ dr $.

Any irreducible unitary representation of $ SO(2) $ is one-dimensional.
It is labeled by an integer $ n \in \Z $ and defined by
\begin{equation}
	\rho_n : SO(2) \to U(1);
	\quad
	\left(
		\begin{array}{rr}
		\cos \alpha & - \sin \alpha \\
		\sin \alpha &   \cos \alpha 
		\end{array}
	\right)
	\mapsto
	e^{i n \alpha}.
	\label{rep of SO(2)}
\end{equation}
Accordingly, an equivariant function $ \psi : \R^2 \to \C $
which satisfies $ \psi (gx) = \rho_n (g) \psi (x) $
becomes a function $ \psi_n (r, \phi) $ 
subject to the condition
$ \psi_n (r, \phi + \alpha) = e^{i n \alpha} \psi_n (r, \phi) $.
Thus we can put $ \psi_n $ in the form 
$ \psi_n (r, \phi) = e^{i n \phi} f_n(r) $.
Here the decomposition (\ref{decomposition}) 
with the projection operators (\ref{projection}) realizes
the ordinary Fourier expansion in angular coordinate,
$ \psi (r, \theta) 
= \sum_{n = - \infty}^{\infty} e^{i n \phi} f_n(r) $.
Since $ SO(2) $ is an isotropy group at the origin $ r=0 $,
smooth equivariant functions $ \psi_n $ must satisfy
\begin{equation}
	\left.
	\frac{\partial}{\partial \phi} \psi_n 
	\right|_{r=0}
	= 0,
	\label{boundary condition of psi}
\end{equation}
which has an alternative expression
\begin{equation}
	n f_n(0) = 0.
	\label{boundary condition of fpsi}
\end{equation}
This boundary condition illustrates the general condition
(\ref{boundary condition3}).

We proceed to reduce the ordinary Laplacian.
The metric on the cotangent bundle $ T^* \R^2 $ is expressed as 
\begin{equation}
	g_M^{-1} =
	\frac{\partial}{\partial r}
	\otimes
	\frac{\partial}{\partial r}
	+
	\frac{1}{r^2}
	\frac{\partial}{\partial \phi}
	\otimes
	\frac{\partial}{\partial \phi}.
	\label{inverse metric}
\end{equation}
To obtain the reduced Laplacian,
we calculate the integral (\ref{Laplacian on equivariant function})
for $ \psi_n $;
\begin{eqnarray}
	\int_{R^2}
	|| d \psi_n ||^2_{g_M} d \mu_M
	& = &
	\int_{R^2}
	\left(
		\left|
		\frac{\partial \psi_n}{\partial r}
		\right|^2
		+
		\frac{1}{r^2}
		\left|
		\frac{\partial \psi_n}{\partial \phi}
		\right|^2
	\right)
	r dr d \phi
	\nonumber \\
	& = &
	\int_0^\infty
	\left(
		\left|
		\frac{d f_n}{d r}
		\right|^2
		+
		\frac{n^2}{r^2}
		\left| 
		f_n
		\right|^2
	\right)
	2 \pi r dr
	\nonumber \\
	& = &
	\left[
		2 \pi r \,
		\overline{f_n} \,
		\frac{d f_n}{d r}
	\right]_0^\infty
	\nonumber \\
	&&
	+ 
	\int_0^\infty
	\overline{f_n}
	\left(
		- \frac{1}{r}
		\frac{d}{d r}
		r
		\frac{d f_n}{d r}
		+
		\frac{n^2}{r^2}
		f_n
	\right)
	2 \pi r dr.
	\label{calculation of Laplacian of R^2}
\end{eqnarray}
Since $ f_n(r) $ is bounded as $ r \to 0 $ from 
(\ref{boundary condition of fpsi}), 
and since $ r\dfrac{d f_n}{d r} \to 0 $ as $ r \to 0 $ 
from the second line of (\ref{calculation of Laplacian of R^2}), 
the boundary term at $ r = 0 $ vanishes.
The other term at $ r = \infty $ vanishes
because of the assumption that 
$ \psi_n $ has a compact support.
Thus we are left with the reduced Laplacian
\begin{equation}
	\Delta_n f_n
	=
	- \frac{1}{r}
	\frac{d}{d r}
	r
	\frac{d f_n}{d r}
	+
	\frac{n^2}{r^2}
	f_n.
	\label{Laplacian of R^2}
\end{equation}
We note here that the boundary condition 
(\ref{boundary condition of fpsi}) says that 
$ f_n(0)=0 $ for $ n \neq 0 $ and that $ f_0(0) $ is 
bounded for $ n=0 $. 
This result gives an example of the general formula 
(\ref{calculated Laplacian}).
An eigenfunction of the Laplacian $ \Delta_n $
with an eigenvalue $ E > 0 $
is the $ n $-th Bessel function $ J_n( \sqrt{E} r ) $.
The Neumann function $ Y_n( \sqrt{E} r ) $ does not
satisfy the boundary condition (\ref{boundary condition of fpsi}).

\subsection{Rigid body}
We have to notice here that our theory covers quantum mechanics of 
rigid bodies.
In mechanics, a rigid body is defined as a collection 
of mass points in three dimensions, in which their mutual 
distances are kept fixed.
In this case, the symmetry group $ G $ is $ SO(3) $ and
the configuration space $ M $ becomes a single $ G $-orbit. 
Hence the orbit space $ Q $ reduces to a single point. 
When the orbit is maximal, $ M $ becomes $ SO(3) $ 
and the inertia tensor $ I $ is nondegenerated. 
When the orbit is singular, $ M $ is isomorphic 
to $ S^2 $ or a single point, 
and the inertia tensor is of rank two or zero, respectively.
For a rigid body, the horizontal component of the Laplacian 
(\ref{Laplacian chi}), or the vibrational energy operator, 
vanishes, and therefore the Laplacian reduces to 
the rotational energy operator;
\begin{equation}
	\Delta^\chi =
	{\mit\Lambda}^\chi = 
	- ( \rho^\chi_* \otimes \rho^\chi_* ) 
	\circ \widetilde{I}^{-1},
	\label{rigid body}
\end{equation}
which is the Casimir operator 
acting on the representation space $ {\cal H}^\chi $
up to a normalization constant. 
In the language of physics, 
since the rigid body executes no vibrational motion, 
it has only rotational energy, 
which is determined by the angular momentum. 
A simple example will be given in Sec. V. 

\section{Quantum molecular mechanics}
\subsection{Jacobi vectors}
In the previous sections, we have set up a general formulation
for reduction of quantum dynamical systems on the configuration 
space $ M $ with symmetry $ G $.
The Hilbert space $ L_2(M) $ is decomposed into 
the orthogonal direct sum of the spaces of equivariant functions
according to the irreducible unitary representations of $ G $,
as was shown in (\ref{decomp on M}). 
The Laplacian $ \Delta_M $ is accordingly reduced to 
the operator $ \Delta^\chi $ of (\ref{Laplacian chi}) 
acting on each space of equivariant functions 
$ L_2(M; {\cal H}^\chi)^G $. 
We have studied quantum mechanics on $ M = \R^2 $ 
with symmetry $ G = SO(2) $ to give a concrete example.
It was a well-known but nontrivial example
in which reduction by symmetry serves 
as the method of separating variables.

Here, we wish to apply the general formulation 
to molecular mechanics, which is the original problem 
that motivates us.
We consider a molecular system consisting of $ N $ atoms in $ \R^3 $.
The configuration of the molecule is described as an ennuple 
$ (\vect{x}_1, \cdots, \vect{x}_N) $ of the position, 
$ \vect{x}_i \in \R^3 $, of each atom. 
Masses of the atoms are denoted by $ (m_1, \cdots, m_N) $
with $ m_i \in \R_{>0} $.
Assume that we are working with the center-of-mass system;
\begin{equation}
	M = 
	\{ ( \vect{x}_1, \cdots, \vect{x}_N ) \in ( \R^3 )^N 
	\, | \, \sum_{i=1}^N m_i \vect{x}_i = 0 \},
	\label{center of mass}
\end{equation}
which is a linear subspace of $ (\R^3)^N $.
Let $ g \in G = SO(3) $ act on 
$ x = (\vect{x}_1, \cdots, \vect{x}_N) \in M $ by
$ gx = (g \vect{x}_1, \cdots, g \vect{x}_N) $.
We call $ M $ and $ Q = M/SO(3) $ 
the {\it molecular configuration space} 
and the {\it shape space}, respectively.
We may regard $ x = (\vect{x}_1, \cdots, \vect{x}_N) $
as a $ 3 \times N $ matrix.
According as the rank of $ x $ is 3, 2, 1, or 0,
the configuration $ x $ is called 
a {\it generic} configuration, 
a {\it planar} one, 
a {\it collinear} one, or 
a {\it collision} one,
respectively.
A generic or planar configuration
has a maximum orbit which is diffeomorphic to $ SO(3) $.
A collinear configuration, 
in which all the atoms are placed along a line,
has a singular orbit which is diffeomorphic to $ S^2 $.
A collision configuration $ x = (0, \cdots, 0) $
has another singular orbit which consists of a single point.
We are going to review a geometric setting on $ M $ in 
what follows. 
The topology of the shape space $ Q $ will be 
studied in the next subsection.

While a tangent vector $ v \in T_x M $ is denoted by
$ v = ( \vect{v}_1, \cdots, \vect{v}_N ) \in (\R^3)^N $
along with the condition $ \sum_{i=1}^N m_i \vect{v}_i = 0 $,
a cotangent vector $ p \in T^*_x M $ is denoted by
$ p = ( \vect{p}_1, \cdots, \vect{p}_N ) \in (\R^3)^N $
along with the condition $ \sum_{i=1}^N \vect{p}_i = 0 $.
The pairing between $ T_xM $ and $ T^*_x M $ is given by
$ \bra p, v \ket := \sum_{i=1}^N ( \vect{p}_i, \vect{v}_i ) $,
where $ ( \cdot, \cdot ) $ denotes 
the standard inner product of $ \R^3 $.
Each element of the Lie algebra $ \xi \in \so $ 
is identified with a vector $ \vect{\xi} \in \R^3 $
and induces the infinitesimal transformation 
$ \theta_x (\vect{\xi}) = 
( \vect{\xi} \times \vect{x}_1, \cdots, 
  \vect{\xi} \times \vect{x}_N ) $,
where $ \times $ means the standard vector product in $ \R^3 $.
Moreover, the dual space of the Lie algebra $ \so^* $ 
is also identified with $ \R^3 $.

The space $ (\R^3)^N $ is equipped with a Riemannian metric $ K $;
\begin{equation}
	K
	:= \sum_{i=1}^N m_i ( d\vect{x}_i, d\vect{x}_i ).
	\label{kinetic energy}
\end{equation}
Although $ K $ is twice the kinetic energy,
we will call it the kinetic energy simply.
The subspace $ M \subset (\R^3)^N $ inherits 
the Riemannian metric $ K $. 
{}From the definition (\ref{angular momentum map}), 
the angular momentum $ \vect{L} : T^* M \to \so^* \cong \R^3 $ 
is expressed as 
\begin{equation}
	( \vect{L}_x(p), \vect{\xi} )
	=
	\bra p, \theta_x (\vect{\xi}) \ket,
	= 
	\sum_{i=1}^N ( \vect{p}_i, \vect{\xi} \times \vect{x}_i )
	= 
	( \sum_{i=1}^N \vect{x}_i \times \vect{p}_i, \vect{\xi} ),
\end{equation}
and hence takes the usual form 
\begin{equation}
	\vect{L} = \sum_{i=1}^N \vect{x}_i \times \vect{p}_i.
	\label{L}
\end{equation}
According to (\ref{dual angular momentum map}),
the angular momentum form $ \widehat{\vect{L}} : TM \to \so^* $ 
is expressed as 
\begin{equation}
	( \widehat{\vect{L}}_x(v), \vect{\xi} )
	=
	K( v, \theta_x(\vect{\xi}) )
	=
	\sum_{i=1}^N m_i ( \vect{v}_i, \vect{\xi} \times \vect{x}_i )
	=
	( \sum_{i=1}^N  m_i \vect{x}_i \times \vect{v}_i, \vect{\xi} ),
\end{equation}
which implies that 
\begin{equation}
	\widehat{\vect{L}} 
	= \sum_{i=1}^N m_i \vect{x}_i \times d\vect{x}_i.
	\label{dual L}
\end{equation}

The Jacobi vectors are of great help in describing 
many-body systems.
Let us remind us of the definition of the Jacobi vectors.
By $ M_i \in \R_{>0} $ and $ \vect{X}_i \in \R^3 $ 
we denote the sum of the masses from the first to 
the $ i $-th atom and the center of mass of the set 
of $ i $ atoms, respectively,
\begin{equation}
	M_i := \sum_{j=1}^i m_j,
	\qquad
	\vect{X}_i := \frac{1}{M_i} \sum_{j=1}^i m_j \vect{x}_j
	\qquad
	(i=1, \cdots, N).
	\label{sum of mass}
\end{equation}
In particular, one has $ \vect{X}_1 = \vect{x}_1 $, 
and $ \vect{X}_N $ is equal to the center of mass of 
the whole system.
Then the Jacobi vectors
$ (\vect{r}_0^{(N)}, \vect{r}_1^{(N)}, \cdots, \vect{r}_{N-1}^{(N)}) $
are defined by
\begin{eqnarray}
	\vect{r}_0^{(N)} 
	& := &
	\sqrt{ M_N } \, \vect{X}_N,
	\nonumber \\
	\vect{r}_i^{(N)} 
	& := &
	\left(
	\frac{1}{M_i} + \frac{1}{m_{i+1}}
	\right)^{ - \frac12}
	( \vect{x}_{i+1} - \vect{X}_i )
	\qquad
	( i = 1, \cdots, N-1 ).
	\label{Jacobi}
\end{eqnarray}
Of course, in the center-of-mass system, one has 
$ \vect{X}_N = \vect{r}_0^{(N)} = 0 $.
The Jacobi vectors 
$ ( \vect{r}_1^{(N)}, \cdots \vect{r}_{N-1}^{(N)} ) $ 
provide a coordinate system to $ M $, and give rise to the isomorphism 
$ M \cong (\R^3)^{(N-1)} $. 

Good use is made of the Jacobi vectors to prove
the additivity of the kinetic energy $ K $
and of the angular momentum $ \widehat{\vect{L}} $ 
in the number of atoms.
In fact, one can verify that
\begin{eqnarray}
	&&
	K^{(N)} 
	:= \sum_{i=1}^N m_i ( d\vect{x}_i, d\vect{x}_i ) 
	 = \sum_{i=0}^{N-1} ( d\vect{r}_i^{(N)}, d\vect{r}_i^{(N)} ),
	\label{additivity of K}
	\\
	&&
	\widehat{\vect{L}}^{(N)} 
	:= \sum_{i=1}^N m_i \vect{x}_i \times d\vect{x}_i 
	 = \sum_{i=0}^{N-1} \vect{r}_i^{(N)} \times d\vect{r}_i^{(N)}.
	\label{additivity of L}
\end{eqnarray}
The additivity of $ K^{(N)} $ can be proved 
by induction with respect to $ N $.
A straightforward calculation yields
\begin{eqnarray}
	&   &
	K^{(N+1)} - K^{(N)}
	\nonumber \\
	& = &
	  || d\vect{r}_0^{(N+1)} ||^2 
	+ || d\vect{r}_N^{(N+1)} ||^2
	- || d\vect{r}_0^{(N)} ||^2
	\nonumber \\
	& = &
	  M_{N+1} || d\vect{X}_{N+1} ||^2 
	+ \frac{ M_N m_{N+1} }{ M_N + m_{N+1} } 
	  || d\vect{x}_{N+1} - d\vect{X}_N ||^2
	- M_N     || d\vect{X}_N ||^2
	\nonumber \\
	& = &
	  \frac{1}{M_{N+1}} || M_N d\vect{X}_N 
	  + m_{N+1} d\vect{x}_{N+1} ||^2 
	+ \frac{ M_N m_{N+1} }{ M_{N+1} } || d\vect{x}_{N+1} 
	- d\vect{X}_N ||^2
	- M_N     || d\vect{X}_N ||^2
	\nonumber \\
	& = &
	\frac{ M_N^2 + M_N m_{N+1} - M_N M_{N+1} }{ M_{N+1} }
	|| d\vect{X}_N ||^2
	+
	\frac{ m_{N+1}^2 + M_N m_{N+1} }{ M_{N+1} }
	|| d\vect{x}_{N+1} ||^2
	\nonumber \\
	& = &
	m_{N+1} || d\vect{x}_{N+1} ||^2.
	\label{proof of additivity of K}
\end{eqnarray}
In a similar manner, 
the additivity of the angular momentum is verified as follows:
\begin{eqnarray}
	&   &
	\widehat{\vect{L}}^{(N+1)} - \widehat{\vect{L}}^{(N)}
	\nonumber \\
	& = &
	  \vect{r}_0^{(N+1)} \times d\vect{r}_0^{(N+1)}
	+ \vect{r}_N^{(N+1)} \times d\vect{r}_N^{(N+1)}
	- \vect{r}_0^{(N)}   \times d\vect{r}_0^{(N)}
	\nonumber \\
	& = &
	  M_{N+1} \vect{X}_{N+1} \times d\vect{X}_{N+1}
	\nonumber \\
	&   &
	\quad
	+ \frac{ M_N m_{N+1} }{ M_N + m_{N+1} } 
	( \vect{x}_{N+1} - \vect{X}_N ) \times ( d\vect{x}_{N+1} 
	- d\vect{X}_N )
	- M_N \vect{X}_N \times d\vect{X}_N
	\nonumber \\
	& = &
	  \frac{1}{M_{N+1}} 
	  ( M_N  \vect{X}_N + m_{N+1}  \vect{x}_{N+1} ) \times 
	  ( M_N d\vect{X}_N + m_{N+1} d\vect{x}_{N+1} )
	\nonumber \\
	&   &
	\quad
	+ \frac{ M_N m_{N+1} }{ M_{N+1} } 
	( \vect{x}_{N+1} - \vect{X}_N ) 
	\times ( d\vect{x}_{N+1} - d\vect{X}_N )
	- M_N \vect{X}_N \times d\vect{X}_N
	\nonumber \\
	& = &
	\frac{ M_N^2 + M_N m_{N+1} - M_N M_{N+1} }{ M_{N+1} }
	\vect{X}_N \times d\vect{X}_N 
	+
	\frac{ m_{N+1}^2 + M_N m_{N+1} }{ M_{N+1} }
	\vect{x}_{N+1} \times d\vect{x}_{N+1}
	\nonumber \\
	& = &
	m_{N+1} \vect{x}_{N+1} \times d\vect{x}_{N+1}.
	\label{proof of additivity of L}
\end{eqnarray}
In the following, we fix the number of atoms $ N $
and suppress the superscript $ {}(N) $.

According to the relations 
(\ref{L theta=I}) and (\ref{additivity of L}),
the inertia operator $ \widehat{I}_x : \so \to \so^* $ is defined 
for each $ \xi \in \so \cong \R^3 $ by
\begin{equation}
	\widehat{I}_x ( \vect{\xi} ) 
	=
	\widehat{\vect{L}} ( \theta_x (\vect{\xi}) )
	=
	\sum_{i=0}^{N-1} \vect{r}_i \times 
	( \vect{\xi} \times \vect{r}_i )
	\nonumber \\
	=
	\sum_{i=0}^{N-1} 
	\bigl(
	( \vect{r}_i, \vect{r}_i ) \, \vect{\xi} 
	- ( \vect{\xi}, \vect{r}_i ) \, \vect{r}_i
	\bigr).
	\label{calculation of inertia tensor}
\end{equation}
According as $ x $ is a generic configuration, a planar one,
a collinear one, or the collision one,
the rank of $ \widehat{I}_x $ is 3, 3, 2, or 0, respectively.
Unfortunately, there is no concise expression 
for the inverse $ ( \widetilde{I}_x )^{-1} $ in general. 
However, the connection form $ \omega $ is expressed, 
from (\ref{calculator of omega}), as 
\begin{equation}
 \omega = \widetilde{I}^{-1}\sum_{i=1}^{N-1}
          \vect{r}_i \times d\vect{r}_i 
      \label{mol conn}
\end{equation}

To formulate molecular mechanics,
we need the invariant volume form $ v_M $ of $ M $
associated to the metric $ K $, 
\begin{equation}
	v_M = d^3 \vect{r}_1 \wedge \cdots \wedge d^3 \vect{r}_{N-1}.
	\label{volume by K}
\end{equation}
Thus we have made a geometric setting to apply our formalism
to molecular mechanics.
However, before application
we have to examine the topology of the orbit space (or shape 
space) $ Q = M/G $
for $ N $-atomic molecules.

\subsection{Topology of the shape space}
Let $ M(m,n) $ and $ M(m,n)_k $
be the vector space of $ m \times n $ matrices over $ \R $,
and
the set of $ m \times n $ matrices of rank $ k $,
respectively.
By $ S(n) $ we denote the set 
of all the positive semi-definite symmetric 
$ n \times n $ matrices,
and set $ S(n)_k := S(n) \cap M(n,n)_k $.
Of course, $ k \le m, n $.
Let $ O(n) $ denote the orthogonal group acting on $ \R^n $ 
as usual.
Then $ O(m) $ and $ O(n) $ act on $ M(m,n) $ to the left and 
to the right, respectively. 
We can verify now that
\begin{equation}
	M(m,n)_k
	\cong
	\frac{
	O(m) / O(m-k) \times S(k)_k \times O(n-k) \backslash O(n)
	}
	{O(k)}
	\label{matrix decomposition}
\end{equation}
from the observation of the fact that 
an arbitrary linear map $ \varphi : \R^n \to \R^m $ of 
rank $ k $ can be 
expressed as a composition $ \varphi = i \circ \sigma \circ \pi $ 
of three linear maps, where
$ \pi $, $ \sigma $, and $ i $ are
an orthogonal submersion $ \pi : \R^n \to \R^k $,
a positive-definite symmetric operator $ \sigma : \R^k \to \R^k $,
and an orthogonal immersion $ i : \R^k \to \R^m $, respectively.
Here we call a linear map 
$ \pi : \R^n \to \R^k $ an orthogonal submersion, 
when it is surjective and satisfies 
$ \pi \circ {}^t \pi = $ id on $ \R^k $.
Similarly, we call $ i : \R^k \to \R^m $ an orthogonal immersion,
when it is injective and satisfies 
$ {}^t i \circ i = $ id on $ \R^k $.
The set of all the orthogonal submersions 
$ \{ \pi : \R^n \to \R^k \} $ is identified with a Stiefel manifold
$ O(n-k) \backslash O(n) $, 
while the set of all the orthogonal immersions
$ \{ i : \R^k \to \R^m \} $ is identified 
with another Stiefel manifold
$ O(m) / O(m-k) $.
An equivalence relation $ \sim $
is defined on the triplet $ ( i, \sigma, \pi ) $
by the action of $ g \in O(k) $ through
$ ( i, \sigma, \pi ) \sim 
( i g^{-1}, g \sigma g^{-1}, g \pi ) $.
In particular, for $ m = n = k $,
Eq.(\ref{matrix decomposition}) becomes
\begin{equation}
	GL(n)
	=
	M(n,n)_n
	\cong
	S(n)_n \times O(n),
	\label{matrix polar decomposition}
\end{equation}
which is nothing but the so-called polar decomposition 
of regular matrices.

Thus the configuration space of the molecule,
$ M = (\R^3)^{N-1} = M(3, N-1) $, is identified with
\begin{equation}
	M(3,N-1)
	\cong
	\bigcup_{0 \le k \le \min(3,N-1)}
	\frac{
	O(3) / O(3-k) \times S(k)_k \times O(N-1-k) \backslash O(N-1)
	}
	{O(k)}.
	\label{decomposition of M}
\end{equation}
Each component with $ k = 0,1,2,3 $ corresponds to the set of
collision, collinear, planar, and generic configurations, 
respectively. 
Note that a point $ x \in M $ is of the maximum type 
or of the singular type, according as 
$ k=2,3 $ or $ k=0,1 $. 
Strata of the shape space $ Q = M/SO(3) $ are then given by
\begin{equation}
	Q^{(N)}_k
	\cong
	\frac{
	SO(3) \backslash O(3) / O(3-k) 
	\times S(k)_k \times O(N-1-k) \backslash O(N-1)
	}
	{O(k)}.
	\label{decomposition of Q}
\end{equation}
The topology of strata for few-body problems with 
$ N = 3 $ and $ N = 4 $ is already studied by 
one of the authors\cite{Iwai3} and
Narasimhan-Ramadas\cite{Narasimhan}, respectively.
Coordinates of the $ N $-body problem are also studied 
by Littlejohn and Reinsch\cite{Littlejohn}.
They also wrote a comprehensive review\cite{Littlejohn2} 
on gauge fields in the $N$-body problem, and studied also 
complexes of rigid molecules\cite{Littlejohn3}. 

We write out topology with $ N=2,3,4 $ to give definite examples.
We denote by $ \R_{>0} $ the positive real numbers $ (0, \infty) $ 
and by $ D^n $, $ S^n $, and $ \R P^n $ 
an $ n $-dimensional disk, sphere, projective space, 
respectively. 
\\
$ N = 2 $:
\begin{eqnarray}
	&&
	Q^{(2)}_1
	\cong
	S(1)_1
	\cong
	\R_{>0},
	\nonumber \\
	&&
	Q^{(2)}_0
	\cong
	\{ 0 \}.
	\label{decomposition of Q^2}
\end{eqnarray}
\\
$ N = 3 $:
\begin{eqnarray}
	&&
	Q^{(3)}_2
	\cong
	S(2)_2
	\cong
	\R_{>0} \times D^2 \cong \R_{>0} \times \R^2,
	\nonumber \\
	&&
	Q^{(3)}_1
	\cong
	S(1)_1 \times 
	\frac{O(2)}{O(1) \times O(1)}
	\cong
	S(2)_1
	\cong
	\R_{>0} \times \R P^1
	\cong
	\R_{>0} \times S^1 \cong \R^2 - \{ 0 \},
	\nonumber \\
	&&
	Q^{(3)}_0
	\cong
	\{ 0 \}.
	\label{decomposition of Q^3}
\end{eqnarray}
\\
$ N = 4 $:
\begin{eqnarray}
	&&
	Q^{(4)}_3
	\cong
	\frac{O(3)}{SO(3)} \times S(3)_3
	\cong
	\Z_2 \times \R_{>0} \times D^5
	\cong
	\R_{>0} \times (S^5 - S^4),
	\nonumber \\
	&&
	Q^{(4)}_2
	\cong
	\frac{S(2)_2 \times O(1) \backslash O(3)}{O(2)}
	\cong
	S(3)_2
	\cong
	\R_{>0} \times (S^4 - \R P^2),
	\nonumber \\
	&&
	Q^{(4)}_1
	\cong
	\frac{S(1)_1 \times O(2) \backslash O(3)}{O(1)}
	\cong
	S(3)_1
	\cong
	\R_{>0} \times \R P^2,
	\nonumber \\
	&&
	Q^{(4)}_0
	\cong
	\{ 0 \}.
	\label{decomposition of Q^4}
\end{eqnarray}
In the case of $ N=3 $, the union of $ Q^{(3)}_k,\,k=0,1,2 $, 
forms the shape space $ Q \cong \R^2 \times \R_{\geq 0} $. 
The maximum stratum is $ Q^{(3)}_2 \cong \R^2 \times \R_{>0} $. 
{}For $ N=4 $, the union $ Q^{(4)}_3 \cup Q^{(4)}_2 $ is 
the maximum stratum, which is diffeomorphic to
$ \R_{>0} \times (S^5 - \R P^2) $.

\subsection{Triatomic molecules}
To make a practical application of the above general formalism,
we concentrate on the triatomic molecules in the rest of the paper.
The configuration space then becomes
$ M = ( \R^3 )^2 = \{ ( \vect{r}_1, \vect{r}_2 ) \} $.
The maximum stratum $ M_{\mu} $ is diffeomorphic with 
$ M(3,2)_2 $, the space of $ 3 \times 2 $ matrices of 
maximal rank. 
The stratum of singular orbit type, $ \partial M_{\mu} $, is 
the union $ M(3,2)_1 \cup M(3,2)_0 $. 
Dragt\cite{Dragt} and his successors 
have introduced a useful coordinate system 
$ ( \alpha, \beta, \gamma, \rho, \chi, \phi ) $ of $ M $ by setting
\begin{eqnarray}
	&&
	\vect{r}_1 = \rho \,
	( \cos \frac{\chi}{2} \cos \frac{\phi}{2} \vect{u}_3
	+ \sin \frac{\chi}{2} \sin \frac{\phi}{2} \vect{u}_2 ),
	\label{r_1}
	\\
	&&
	\vect{r}_2 = \rho \,
	( \cos \frac{\chi}{2} \sin \frac{\phi}{2} \vect{u}_3
	- \sin \frac{\chi}{2} \cos \frac{\phi}{2} \vect{u}_2 ).
	\label{r_2}
\end{eqnarray}
Here $ ( \vect{u}_1, \vect{u}_2, \vect{u}_3 ) $ 
is an orthonormal basis of $ \R^3 $ 
parametrized by the Euler angles $ ( \alpha, \beta, \gamma ) $ as
\begin{equation}
	  ( \vect{u}_1, \vect{u}_2, \vect{u}_3 ) 
	= g( \vect{e}_1, \vect{e}_2, \vect{e}_3 ),
\quad 
    g=e^{\alpha J_3} e^{\beta J_2} e^{\gamma J_3}, 
	\label{Euler angle}
\end{equation}
where $ ( \vect{e}_1, \vect{e}_2, \vect{e}_3 ) $ 
is the standard basis of $ \R^3 $ and 
$ g $ is an element of $ SO(3) $ with 
$ ( J_1, J_2, J_3 ) $ the standard basis of $ \so $
defined by 
$ J_i \vect{v} = \vect{e}_i \times \vect{v} $ 
$ (i=1,2,3) $ for each $ \vect{v} \in \R^3 $.
We notice here that the orientation of the frame 
$ (\vect{u}_1, \vect{u}_2, \vect{u}_3 ) $
is different from that of the original article. 
We choose the orientation to bring the collinear configurations
into the direction of $ \vect{u}_3 $. 
The ranges of the coordinates are given by
\begin{eqnarray}
	&&
	0 \le \alpha <   2 \pi, \quad
	0 \le \beta  \le   \pi, \quad
	0 \le \gamma \le   2 \pi, 
	\nonumber \\
	&&
	0 \le \rho,                   \quad
	0 \le \chi \le \frac{\pi}{2}, \quad
	0 \le \phi <       2 \pi.
	\label{range}
\end{eqnarray}
The geometric meaning of $ (\rho, \chi, \phi) $ is clear
on introducing coordinates $ (q_1, q_2, q_3 ) $ by
\begin{eqnarray}
	&&
	q_1 
	:= || \vect{r}_1 ||^2 - || \vect{r}_2 ||^2
	 = \rho^2 \, \cos \chi \cos \phi,
	\label{q_1} \\
	&&
	q_2 
	:= 2 ( \vect{r}_1, \vect{r}_2 )
	 = \rho^2 \, \cos \chi \sin \phi,
	\label{q_2} \\
	&&
	q_3 
	:= 2 || \vect{r}_1 \times \vect{r}_2 ||
	 = \rho^2 \, \sin \chi.
	\label{q_3}
\end{eqnarray}
They are invariant under the action of $ SO(3) $ on $ M $,
and provide the projection $ \pi:\, M \to Q = M/SO(3) $; 
$ (\vect{r}_1,\vect{r}_2) \mapsto (q_1,q_2,q_3) $.  
With this expression of $ \pi $, it is easy to show that
$ Q $ is homeomorphic to the upper half space 
$ \Rplus^3 = \R^2 \times \Rplus $. 
The space, $ \pi(M_{\mu}) $, of maximum orbits is diffeomorphic 
with $ \R^2 \times \R_{>0} $.  
The boundary surface, $ \R^2 \times \{0\} $, determined by 
$ q_3 = 0 $ or $ \chi = 0 $ describes the orbit space for 
collinear configurations,
in which the atoms make a line along $ \vect{u}_3 $.
The origin $ (0,0,0) $ represents the collision configuration.
These observations coincide with (\ref{decomposition of Q^3}).
On the other hand, the Euler angles $ ( \alpha, \beta, \gamma ) $
are regarded
as a coordinate system of the fiber of the bundle
$ \pi : M_\mu \to Q_\mu $.
Note that,
in the set of singular points $ \partial M_\mu $, 
one has 
$ \vect{r}_1 = \rho \cos \dfrac{\phi}{2} \, \vect{u}_3 $,
$ \vect{r}_2 = \rho \sin \dfrac{\phi}{2} \, \vect{u}_3 $ 
with $ \vect{u}_3 = e^{\alpha J_3} e^{\beta J_2} \vect{e}_3 $, 
which shows that $ ( \alpha, \beta ) $ serve as coordinates
for the orbit diffeomorphic with $ S^2 $,
and that $ ( \rho, \phi ) $ are coordinates for 
$ \partial Q_\mu \cong \R^2 $.

In terms of the coordinates introduced above,
we are to write out the geometric objects in the explicit form.
It is also convenient for later use
to introduce the Maurer-Cartan one-form;
\begin{eqnarray}
	&&
	g^{-1} dg
	\nonumber \\
	& = &
	J_1 
	( \sin\gamma \, d\beta - \sin\beta \cos\gamma \, d\alpha )+
	J_2 
	( \cos\gamma \, d\beta + \sin\beta \sin\gamma \, d\alpha )+
	J_3 
	(            \, d \gamma+ \cos \beta          \, d\alpha ) 
	\nonumber \\
	& = &
	J_1 \, {\mit\Theta}_1 +
	J_2 \, {\mit\Theta}_2 +
	J_3 \, {\mit\Theta}_3.
	\label{Maurer-Cartan}
\end{eqnarray}
Then the metric $ g_M = K $ 
obtained in (\ref{additivity of K}) takes the form
\begin{eqnarray}
	g_M
	& = &
	|| d \vect{r}_1 ||^2 + || d \vect{r}_2 ||^2
	\nonumber \\
	& = &
	d \rho^2 
	+ \frac14 \rho^2 \, d \chi^2 
	+ \frac14 \rho^2 \cos^2 \chi \, d \phi^2
	\nonumber \\
	&&
	+ \rho^2 ( {\mit\Theta}_1 - \frac12 \sin \chi \, d \phi )^2
	+ \rho^2 \cos^2 \frac{\chi}{2} \, {\mit\Theta}_2^2
	+ \rho^2 \sin^2 \frac{\chi}{2} \, {\mit\Theta}_3^2.
	\label{calculation of K}
\end{eqnarray}
The angular momentum form 
$ \widehat{\vect{L}} : TM \to \so^* \cong \R^3 $
obtained in (\ref{additivity of L}) is expressed as
\begin{eqnarray}
	\widehat{\vect{L}}
	& = & 
	\vect{r}_1 \times d\vect{r}_1 + \vect{r}_2 \times d\vect{r}_2
	\nonumber \\
	& = &
	\rho^2 \, \vect{u}_1 
	( {\mit\Theta}_1-\frac12 \sin \chi \, d \phi ) +
	\rho^2 \cos^2 \frac{\chi}{2} \, \vect{u}_2 \, {\mit\Theta}_2 +
	\rho^2 \sin^2 \frac{\chi}{2} \, \vect{u}_3 \, {\mit\Theta}_3. 
	\label{calculation of L}
\end{eqnarray}
The inertia operator 
$ \widehat{I}_x : \so \cong \R^3 \to \so^* \cong \R^3 $ 
obtained in (\ref{calculation of inertia tensor}) is put in the form
\begin{eqnarray}
	\widehat{I}_x ( \vect{\xi} )
	& = & 
	\vect{r}_1 \times ( \vect{\xi} \times \vect{r}_1 ) + 
	\vect{r}_2 \times ( \vect{\xi} \times \vect{r}_2 )
	\nonumber \\
	& = &
	\rho^2                       
	( \vect{\xi}, \vect{u}_1 ) \, \vect{u}_1 +
	\rho^2 \cos^2 \frac{\chi}{2} 
	( \vect{\xi}, \vect{u}_2 ) \, \vect{u}_2 +
	\rho^2 \sin^2 \frac{\chi}{2} 
	( \vect{\xi}, \vect{u}_3 ) \, \vect{u}_3. 
	\label{calculation of I}
\end{eqnarray}
The connection form $ \vect{\omega} : TM \to \so \cong \R^3 $
is then written out, according to (\ref{calculator of omega}), as
\begin{equation}
	\vect{\omega}
	= 
	\widetilde{I}^{-1} \widehat{\vect{L}}
	= 
	\vect{u}_1 ( {\mit\Theta}_1 - \frac12 \sin \chi \, d \phi ) +
	\vect{u}_2 \, {\mit\Theta}_2 +
	\vect{u}_3 \, {\mit\Theta}_3. 
	\label{calculation of omega}
\end{equation}
Hence the vertical component of the metric (\ref{rotational energy}) 
is given by
\begin{eqnarray}
	g_M|V
	& = & 
	I \circ ( \vect{\omega} \otimes \vect{\omega} )
	= ( \widehat{\vect{L}}, \vect{\omega} )
	\nonumber \\
	& = &
	  \rho^2 ( {\mit\Theta}_1 - \frac12 \sin \chi \, d \phi )^2
	+ \rho^2 \cos^2 \frac{\chi}{2} \, {\mit\Theta}_2^2
	+ \rho^2 \sin^2 \frac{\chi}{2} \, {\mit\Theta}_3^2
\end{eqnarray}
and thereby the metric $ g_Q $ such that $ \pi^* g_Q = g_M|H $ 
becomes 
\begin{equation}
	g_Q
	= 
	d \rho^2 
	+ \frac14 \rho^2 \, d \chi^2 
	+ \frac14 \rho^2 \cos^2 \chi \, d \phi^2.
	\label{calculation of projected K}
\end{equation}
Then its inverse is a tensor field given by
\begin{equation}
	(g_Q)^{-1}
	= 
	\frac{\partial}{\partial \rho}
	\otimes
	\frac{\partial}{\partial \rho}
	+ 
	\frac{4}{\rho^2}
	\frac{\partial}{\partial \chi}
	\otimes
	\frac{\partial}{\partial \chi}
	+
	\frac{4}{\rho^2 \cos^2 \chi}
	\frac{\partial}{\partial \phi}
	\otimes
	\frac{\partial}{\partial \phi}.
	\label{calculation of inverse of K}
\end{equation}
The volume form defined by (\ref{calculation of K}) is of the form
\begin{equation}
	v_M 
	=
	\frac{1}{16} \rho^5 \sin 2 \chi \,
	d \rho 
	\wedge d \chi 
	\wedge d \phi 
	\wedge {\mit\Theta}_1
	\wedge {\mit\Theta}_2
	\wedge {\mit\Theta}_3.
	\label{calculation of the volume by K}
\end{equation}
Then the volume form associated with the measure (\ref{muQ}) is
\begin{equation}
	v_Q
	=
	\frac{\pi^2}{2} \rho^5 \sin 2 \chi \,
	d \rho 
	\wedge d \chi 
	\wedge d \phi,
	\label{calculation of the push-forwarded measure}
\end{equation}
since 
$ \int {\mit\Theta}_1 
\wedge {\mit\Theta}_2 
\wedge {\mit\Theta}_3 = 8 \pi^2 $.

To describe quantum mechanics for the triatomic molecule,
we need the Hilbert space of sections
of vector bundles associated with the stratification.
Any irreducible unitary representation of $ SO(3) $ is
characterized by a nonnegative integer $ \ell $ and
denoted by $ \rho^\ell : SO(3) \to U( \C^{2 \ell + 1} ) $.
We put $ \widehat{J}_i := (\rho^\ell)_* (\vect{e}_i) $ for 
the standard basis $ \vect{e}_i \in \R^3 \cong \so $.
Since the base space of the stratified fiber bundle
$ \pi : M \to Q \cong \R^2 \times \R_{\ge 0} $ is contractible, 
the bundle is a trivial bundle.
Through a global section $ \sigma : Q \to M $
with $ \alpha = \beta = \gamma = 0 $
in (\ref{r_1}), (\ref{r_2}), and (\ref{Euler angle}),
any equivariant function $ \psi : M \to \C^{2 \ell + 1} $ 
is pulled back to a function 
$ {\mit\Psi} := \sigma^* \psi : \Rplus^3 \to \C^{2 \ell + 1} $. 
Then the boundary condition (\ref{boundary condition3}) takes 
the form 
\begin{equation}
	\widehat{J}_3 \, {\mit\Psi} = 0
	\quad \mbox{for} \quad 
	\chi = 0
	\label{vanishing condition 1}
\end{equation}
on $ M(3,2)_1 $ and
\begin{equation}
	\widehat{J}_1 \, {\mit\Psi} = 
	\widehat{J}_2 \, {\mit\Psi} = 
	\widehat{J}_3 \, {\mit\Psi} = 0
	\quad \mbox{for} \quad 
	\rho = 0
	\label{vanishing condition 2}
\end{equation}
on $ M(3,2)_0$, respectively, with the coordinate system 
defined at (\ref{q_1}), (\ref{q_2}), and (\ref{q_3}).
What Eq. (\ref{vanishing condition 1}) means is as follows: 
At a collinear configuration determined by $ \chi = 0 $,
the molecule lying in the line along $ \vect{u}_3 $ 
has the vanishing angular momentum about $ \vect{u}_3 $; 
$ (\rho^{\ell})_* (\vect{u}_3)\psi(x)=0 $, so that one has 
$ \rho^{\ell}(g)(\rho^{\ell})_*(\vect{e}_3)\rho^{\ell}(g^{-1})
\psi(g \sigma(q))=0 $ with $ \vect{u}_3 =g \vect{e}_3 $, 
which is equivalent to (\ref{vanishing condition 1}). 
Since 
$ \widehat{J}_3=\mbox{diag}(\ell,\ell -1,\dots,0,\dots,-\ell)$, 
the components $ \mit\Psi_m $ of $ \mit\Psi $ with $ m\neq 0 $ 
vanish, if $ \ell \neq 0 $. 
{}Furthermore, Eq. (\ref{vanishing condition 2}) means that
at the collision configuration determined by $ \rho = 0 $,
the molecule cannot carry nonzero angular momentum,
so that the wave function can have a nonvanishing value
only when $ \ell = 0 $.
These conditions are analogs of that for the two-dimensional case
(\ref{boundary condition of fpsi}).
Now we have implemented
the consideration of singular case by providing
the boundary condition (\ref{vanishing condition 1})
and (\ref{vanishing condition 2})
which we skipped in the previous work\cite{Iwai3}.

In conclusion, we write down the Laplacian 
(\ref{calculated Laplacian}), combining (\ref{calculation of I}),
(\ref{calculation of omega}),
(\ref{calculation of inverse of K}), and
(\ref{calculation of the push-forwarded measure});
\begin{eqnarray}
	&&
	- \Delta {\mit\Psi} ( \rho, \chi, \phi )
	\nonumber \\
	& = &
	\frac{1}{\rho^5 \sin 2 \chi}
	\left\{
	\frac{\partial}{\partial \rho}
	\rho^5 \sin 2 \chi
	\frac{\partial}{\partial \rho}
	+
	\frac{\partial}{\partial \chi}
	\frac{4}{\rho^2}
	\rho^5 \sin 2 \chi
	\frac{\partial}{\partial \chi}
	\right.
	\nonumber \\
	&&
	\qquad \qquad
	\left.
	+
	\left( 
		\frac{\partial}{\partial \phi} 
		+ \frac12 \sin \chi \widehat{J}_1
	\right)
	\frac{4}{\rho^2 \cos^2 \chi}
	\rho^5 \sin 2 \chi
	\left( 
		\frac{\partial}{\partial \phi} 
		+ \frac12 \sin \chi \widehat{J}_1
	\right)
	\right\}
	{\mit\Psi}
	\nonumber \\
	&&
	\qquad
	+
	\frac{1}{\rho^2}
	\left\{
	(\widehat{J}_1)^2
	+ \frac{1}{\cos^2 ( \chi/2 ) }
	(\widehat{J}_2)^2
	+ \frac{1}{\sin^2 ( \chi/2 ) }
	(\widehat{J}_3)^2
	\right\}
	{\mit\Psi}
	\nonumber \\
	& = &
	\left\{
	\frac{\partial^2}{\partial \rho^2}
	+
	\frac{5}{\rho}
	\frac{\partial}{\partial \rho}
	+
	\frac{4}{\rho^2}
	\left(
		\frac{\partial^2}{\partial \chi^2}
		+
		2 \cot 2 \chi
		\frac{\partial}{\partial \chi}
	\right)
	+
	\frac{4}{\rho^2 \cos^2 \chi}
	\left( 
		\frac{\partial}{\partial \phi} 
		+ \frac12 \sin \chi \widehat{J}_1
	\right)^2
	\right\}
	{\mit\Psi}
	\nonumber \\
	&&
	\qquad
	+
	\frac{1}{\rho^2}
	\left\{
	(\widehat{J}_1)^2
	+ \frac{1}{\cos^2 ( \chi/2 ) }
	(\widehat{J}_2)^2
	+ \frac{1}{\sin^2 ( \chi/2 ) }
	(\widehat{J}_3)^2
	\right\}
	{\mit\Psi},
	\label{final result}
\end{eqnarray}
which reproduces the result of \cite{Iwai3}.
The first and last terms of the RHS of (\ref{final result}) 
are vibrational and rotational energy operators, 
up to sign, respectively. 
As was pointed out in \cite{Iwai3}, if the vibrational energy 
operator is separated off, and if the internal coordinates 
$(\rho,\chi,\phi)$ are fixed, 
the operator $ \frac12 \Delta $ reduces 
to the well known Hamiltonian for a rigid rotor of plane body. 

\section{Concluding remarks}
In this paper we formulated the general method of reduction of
quantum systems with symmetry by the use of the Peter-Weyl theorem.
Although the method is well-known implicitly among Physicists,
we developed it explicitly to give rigorous grounds 
to quantum mechanics describing molecular motions. 
We studied the stratification of manifolds according to the action 
of a symmetry Lie group and 
then defined a stratified bundle and a stratified connection
as generalization of fiber bundles and connections.
Further, we showed that the reduced quantum system is a pair of 
the Hilbert space and the Hamiltonian 
which are the space of equivariant functions
and the Laplacian expressed in terms of covariant derivation 
with the stratified connection, respectively.
We found the boundary condition that is imposed on
the equivariant functions to make the reduced Laplacian 
a self-adjoint operator.
{}Finally, the general formulation for reduction was applied to
$ N $-atomic molecules, and triatomic molecules were examined 
in particular.

The stratified connection on the stratified bundle
is newly introduced
as a generalization of
connections on principal fiber bundles
and is used to describe the reduced Laplacian.
One of our main results is to have determined the boundary 
condition for making the reduced Laplacian self-adjoint. 
Emmrich and H. R{\"o}mer\cite{Emmrich} analyzed Laplacians 
on orbifolds to study quantization of systems with gauge symmetry.
They found that 
the Laplacian on an orbifold is not essentially self-adjoint
and therefore its self-adjoint extension is not unique.
According to our method, the reduced Laplacian is self-adjoint 
by its definition and 
the boundary condition is accompanied automatically by the symmetry.

We would like to make some comments on remaining problems.
{}First, although we built a general formulation to deal
with quantum molecular systems, we do not obtain spectrum
of the reduced Hamiltonian.
Even the three-body problem is difficult to solve. 
It is desired to develop an approximate method
to solve the reduced eigenvalue problem 
of physically interesting systems.

Second, for application to real molecules,
electronic structure must be considered.
Of course, spins and statistics of electrons and nucleus also
must be taken into account.
These are left for future investigation.

\section*{Acknowledgment}
The authors would like to thank Professor Uwano for helpful discussion.
They are also grateful to Professor Tachibana 
for bringing our attention to problems of molecular quantum mechanics.
This work was supported by a Grant-in-Aid for Scientific Research from
the Ministry of Education, Science and Culture of Japan.

\baselineskip 5mm 

\end{document}